\documentclass[a4paper,12pt]{article}

\usepackage[utf8]{inputenc}
\usepackage[article]{mymath}
\usepackage[nottoc]{tocbibind}
\usepackage[english]{babel}
\usepackage{marginnote}
\usepackage{enumitem}

\numberwithin{equation}{subsection}
\reversemarginpar

\title{On the relativistic theory of force-free particles with any spin\\
{\normalsize (Über die relativistische Theorie kräftefreier Teilchen mit beliebigem Spin)\\ Helvetica Physica Acta 12, I, 3-37, 1939}}

\author{Markus Fierz\\{\small(translated by Guglielmo Pasa\footnote{guglielmo.pasa@gmail.com}\ )}}

\date{1939\\{\small(\today)}}
\newcommand{\tocadd}[1]{\addcontentsline{toc}{section}{\protect\numberline{}#1}}
\newcommand{\subtocadd}[1]{\addcontentsline{toc}{subsection}{\protect\numberline{}#1}}
\renewcommand\theequation{\arabic{subsection}.\arabic{equation}}

\newcommand{\dt}{\dot{\tau}}
\newcommand{\dr}{\dot{\rho}}
\newcommand{\dl}{\dot{\lambda}}
\newcommand{\dn}{\dot{\nu}}
\newcommand{\dm}{\dot{\mu}}
\newcommand{\bx}{\mathbf{x}}
\newcommand{\bk}{\mathbf{k}}

\newcommand{\Note}[1]{\footnote{{\color{blue}TN: {#1}}}}
\newcommand{\TN}[1]{\Note{ {#1}}}

\begin{document}
\maketitle
\tableofcontents
\pagebreak
\marginnote{p.3}

\begin{abstract}
In the free case, it is possible to define quantum fields which describe particles with integer or half-integer spin larger than one. It is shown that particles with integer spin must have Bose statistic and particles with half-integer-spin must follow Fermi-Dirac statistic. In the free case, the fields with spin smaller or equal to one are already well defined, so that for them alone, the charge and energy density are clearly identified and gauge invariant; while for the higher spin this is the case only for the total charge and total energy. 
\end{abstract}

\section*{Introduction}
\tocadd{Introduction}%
In the present paper we will investigate the relativistic field theory of particles with any, but constant, integer or half-integer spin, which can then be quantized following Jordan and Pauli \cite{jordan:pauli}.
We have restricted ourself, for the time being, to the force-free case. It happens that by giving the spin and mass of the particle that the corresponding field is already clearly defined; moreover the statistic of the particle is fixed by the spin.

From the very general demand that the commutation relation of the fields must be relativistically invariant and infinitesimal and that the energy must be positive definite, follows that particles with integer spin must always obey Bose statistic, and particles with half-integer spin Fermi-Dirac statistic.

While it is possible to describe fields with integer spin with tensors, whereby one can circumvent the spinor calculus of van der Waerden \cite{waerden}, in the case of half-integer spin it didn't seem possible to replace that quite difficult calculus by something more transparent. However, this technical difficulty is probably related to the physical circumstances that the fields with half-integer spin can never correspond to classical fields, because of the exclusion principle, and also never be observable in the classical sense, in contrast to the tensor fields satisfying the Bose statistic.

\marginnote{p.4}
The wave equations considered here were essentially already given by Dirac \cite{dirac}. However the physical meaning of these 
equations in Dirac's work is not clear, there are often informations that can lead to misunderstandings, or are inaccurate.
Especially so, the $H$-operator of Dirac, that he designate as the ``hamiltonian'' or ``energy operator'', is completely 
different from the field energy in the usual meaning. Moreover we cannot simply describe the interaction with the electric 
fields as Dirac claimed, by replacing $\vec{p}$ with $\vec{p}-\frac ec\vec{\phi}$, since the compatibility of the field 
equations will be broken. 
Even the remark that in general the spin is characterized by two numbers $k$ and $l$ doesn't seem to meet with the physical 
facts, that only the sum $k+l$ is important, and the division in the summands $k$ and $l$ seem to be only formal.

Sakata and Yukawa \cite{sakata:yukawa} have also published a note on these Dirac equalities, that seem unfortunate to us.
What is designated as a current density is in general not a vector, but a part of a higher level tensor. Especially in the 
case of spin 1 ($k=1$, $l=\frac12$) these quantities are the energy and momentum densities, that is the $k4$-components of the 
energy-momentum tensor, as we can immediately recognize by comparison with the Maxwell theory. Furthermore, these authors also 
investigate the spin values which belong the the equations given by Dirac.However, they seem to be too formal in this case, since the representations of the rotation group, according to which the field magnitudes can be transformed, can be considered. This is, however, only sufficient if one is sure that the orbital momentum is zero, e.g. In the rest frame of a plane wave. In this case, it is found that because of the wave equations, the field magnitudes are not only irreducible to Lorentz transformations but also irreducible to rotations. In the rest frame of a wave vector, there are just $2k+2l$ linearly independent plane waves which are transformed with each other in the case of rotations according to the representation $\theta_{k+l-\frac12}$. Since all equations with fixed $k+l$ are mathematically equivalent, it follows that the spin has the value $k+l-\frac12$.

In a diploma thesis, for the first time, Mr. Jauch has given the right expressions for energy-momentum tensors and current vectors, which can be assigned to the Dirac fields. 
\marginnote{p.5}
The special case of spin 1 was dealt with in detail by various authors, especially with regard to nuclear forces. The c-number theory was first considered by Proca, the four-dimensional commutation
relations for this case has been written by Stückelberg \cite{stuckelberg}, and in another form by Kemmer \cite{kemmer}.

Although the present study shows that at least in force-free case fields with any half-integer or integer spin are possible, the small spin values $0,\frac12, 1$ are in some respects excellent.
In these three cases, both energy and charge density are already unambiguously determined in the force-free case, and in the case of 0 and 1 the energy density is positive definite, and in the case of 1/2 the charge density of the c-number theory is positive definite. This is not anymore the case, if the spin becomes lager than 1, only the overall energy or the overall charge are then still clearly defined.
Further, in the quantum theory for spin larger than 1, the charge density on different positions but same time do not commute, rather the derivatives of the $D(x)$-function remains. There are further complications in the attempt to introduce interactions with other fields, even in c-number theory, if the spin is larger than 1. This last point therefore still requires a thorough investigation.

\section{Integer spin}
\subsection{Field tensors and wave equations}
We want to consider force-free, classical wave fields, which can be assigned by means of the relativistic field quantization according to Jordan and Pauli, particles with mass m and spin $\hbar f$. $f$ must be a positive integer.
Such a wave field can be described in the force-free case by a, generally complex, symmetric world tensor $A_{ik\dots l}$ of the $f$-th rank ($f$ indices), which satisfies the wave equation
\begin{equation}\label{eq1.1}
\square A_{ik\dots l}=\kappa^2 A_{ik\dots l}
\end{equation}
where
$$
\square\equiv\sum_{i=1}^4\frac{\partial^2}{\partial x_i^2};\quad (x_i)=(x,y,z,ict)
$$
$\kappa$ is a constant of dimension the inverse of a length, which define a characteristic frequency of the field.
The field quanta assigned to the field are given the mass
$$
\frac{\hbar\kappa}{c}=m.
$$

\marginnote{p.6}
Further $A_{ik\dots l}$ satisfies the two secondary conditions
\begin{alignat}{1}
A_{ii\dots l}&=0\label{eq1.2}\\
\nonumber\\ 
\frac{\partial A_{ik\dots l}}{\partial x_i}&=0\label{eq1.3}
\end{alignat}
The physical meaning of these equations is as follows: The wave equation (\ref{eq1.1}), which is of the Schrödinger-Gordon's type has the consequence that the relativistic theory of classical mass points is contained as a limiting 
case in the quantized field theory.

The secondary conditions (\ref{eq1.2}) and (\ref{eq1.3}) primarily ensure that only particles with the spin $f$ and not 
even those of smaller spins can be assigned to the tensor field. Out of
(\ref{eq1.2}), because of the symmetry of $A_{ik\dots l}$, all the traces of the field tensor disappear.
As shown in the appendix, therefore, $A_{ik\dots l}$ has $(f + 1)^2$ linearly independent components, which transform with 
Lorentz transformations in accordance with the irreducible representation $\theta_{\frac f2,\frac f2}$ of the Lorentz group.
If the fields $A_{ik\dots l}$ is to belong to the spin $f$, then for each given wave number and frequency $k_i$ which satisfy 
the equation  $k_ik_i = -\kappa^2$, there must be $2f+1$ linearly independent plane waves, which differ by the orientation of 
the spin.
That this is the case, we see as follows: Consider the plane waves in question in their rest frame, which always exists if 
$\kappa\not=0$ and $k_4=\pm i\kappa$. There the secondary condition (\ref{eq1.3}) says that all components of $A_{ik\dots l}$, 
in which one of the indices is equal to 4, vanish; therefore, in the rest frame, the indices run only from 1 to 3, and 
$A_{ik\dots l}$ has the form
$$
A_{ik\dots l}=A^0_{ik\dots l}e^{\kappa x_4}.
$$
$A^0_{ik\dots l}$ is a symmetric, spatially constant tensor of the $f$-th rank in $\cR_3$ whose traces vanish. A tensor of 
this kind has $2f+1$ linearly independent components, which are transformed into one another by rotations of the coordinate 
system after the irreducible representation $\theta_f$ of the rotation group. It follows that the corresponding particle 
states differ by the $2f+1$ different orientations of the spin. Furthermore, the secondary condition (\ref{eq1.3}) allows the 
establishment of an energy-momentum tensor for the $A$-field and ensures that the overall energy remains positive, which makes 
possible a physical interpretation of the theory.
\marginnote{p.7}

In the discussion of the secondary conditions (\ref{eq1.3}), we made use of the existence of a rest frame for each plane wave. 
The existence of the rest frame is essential for the entire conclusion, and the case $m=0$ must therefore be regarded as a 
degenerate boundary case. We shall, therefore, regard the case of vanishing mass as a separate one. In this case, let us only 
mention that, in this case, there are only two physically truly different plane waves of the same wave number and frequency, 
as is known from the electromagnetic waves which correspond to $f=1$.

We can now replace the differential equations of the $A$-field with a first-order equation-system, which will be convenient 
for the following considerations. The equations are:
\begin{alignat}{1}
B^{(1)}_{[ik]r\dots l}&=\frac{\partial A_{kr\dots l}}{\partial x_i}-\frac{\partial A_{ir\dots l}}{\partial x_k}\label{eq1.4}\\
\nonumber\\
\frac{\partial}{\partial x_i}B^{(1)}_{[ik]r\dots l}&=\kappa^2 A_{kr\dots l}.\label{eq1.5}
\end{alignat}
These equations are analogous to Maxwell's equations. As a consequence of (\ref{eq1.2}), (\ref{eq1.4}) and (\ref{eq1.5}),
 $B^{(1)}$ also satisfies the second-order wave equation, as well as the other equations
\begin{alignat}{1}
B^{(1)}_{[ik]k\dots}=0,&\quad 
B^{(1)}_{[ik]r\dots l}+B^{(1)}_{[ri]k\dots l}+B^{(1)}_{[kr]i\dots l}=0\label{eq1.6}\\
\nonumber\\
\frac{\partial}{\partial x_r}B^{(1)}_{[ik]r\dots l}=0,&\quad
\frac{\partial}{\partial x_m}B^{(1)}_{[ik]\dots l}+\frac{\partial}{\partial x_k}B^{(1)}_{[mi]\dots l}
+\frac{\partial}{\partial x_i}B^{(1)}_{[km]\dots l}=0.\label{eq1.7}
\end{alignat}
In the unbracketed indices, $B_{[ik]r\dots l}$ is symmetric, and skewed in the bracketed pair. From Equations (\ref{eq1.6})
it follows, independently of the definition of $B^{(1)}_{[ik]r\dots l}$, that $B^{(1)}_{[ik]rr\dots l} = 0$. If one takes this 
into account, one can count the number of components of $B^{(1)}$ and find that $B^{(1)}$ has $2f^2+4f$ linearly independent 
components. From the equations (\ref{eq1.6}), (\ref{eq1.7}), all the other equations can be inferred, so that we can describe 
our wave field just as well by the quantities $B^{(1)}_{[ik]r\dots l}$ as by the $A_{ik\dots l}$. We can now form another 
quantity from $B^{(1)}$
$$
B^{(2)}_{[ik][rl]\dots t}
$$
such that
$$
B^{(2)}_{[ik][rl]\dots t}
=\frac{\partial}{\partial x_r}B^{(1)}_{[ik]l\dots t}-\frac{\partial}{\partial x_l}B^{(1)}_{[ik]r\dots t}.
$$
\marginnote{p.8}
$B^{(2)}$ is symmetric when the index pairs $[i k]$ and $[r l]$ are exchanged, as well as in the unbracketed indices, and it 
satisfies analogous equations such as $B^{(1)}$. Further, 
$$
B^{(2)}_{[kl][kr]\dots s} = \kappa^2 A_{lr\dots s}.
$$
By continuing to form the rotation after one of the unbracketed indices, we obtain a sequence of $f + 1$ field quantities:
$$
 A_{ik\dots l},\ B^{(1)}_{[ik]\dots r},\ \dots,\ B^{(f)}_{[ik][rl]\dots[st]}
$$
$B^{(q)}_{[ik]\dots[rl]s\dots t}$ accordingly contains $(q)$ brackets, which are symmetrical to one another and $f - q$ are 
symmetric single indices. In each bracketed index pair, $B^{(q)}$ is skewed. Further, $B^{(q)}$ satisfies the following 
equations:
\begin{alignat}{1}
\square B^{(q)}&=\kappa^2B^{(q)}\tag{I}\label{I}\\
\nonumber\\
B^{(q)}_{\dots[ik]k\dots}&=0\tag{IIa}\label{IIa}\\
\nonumber\\
B^{(q)}_{\dots[ik]r\dots}+B^{(q)}_{\dots[ri]k\dots}+B^{(q)}_{\dots[kr]i\dots}&=0\tag{IIb}\label{IIb}\\
B^{(q)}_{[ik][rs]\dots}+B^{(q)}_{\dots[ri][ks]\dots}+B^{(q)}_{[kr][is]\dots}&=0\nonumber\\
\nonumber\\
\frac{\partial}{\partial x_r}B^{(q)}_{[ik]\dots r}=0;\quad
\frac{\partial}{\partial x_r}B^{(q)}_{\dots[ik]\dots}+& \frac{\partial}{\partial x_k}B^{(q)}_{\dots[ri]\dots}+\frac{\partial}{\partial x_i}B^{(q)}_{\dots(kr)\dots}=0.\tag{III} \label{III}
\end{alignat}
These equations describe the observed wave field exactly as the equations (\ref{eq1.1}) to (\ref{eq1.3}).

It applies as a consequence from \ref{I} to \ref{III}:
$$
\begin{aligned}
\frac{\partial}{\partial x_i}B^{(q)}_{\dots[ik]\dots}=\kappa^2 B^{(q-1)}_{\dots k\dots}\\
\\
\frac{\partial}{\partial x_i}B^{(q-1)}_{k\dots}-\frac{\partial}{\partial x_k}B^{(q-1)}_{\dots i\dots}=B^{(q)}_{\dots[ik]}\\
\\
B^{(q)}_{[kl][ki]\dots}=\kappa^2 B^{(q-2)}_{\dots li\dots}.
\end{aligned}
$$

\subsection{Energy-momentum tensor, current vector}
In order to be able to interpret the tensor field described by $B^{(q)}$, it must be possible to form a symmetric "real" 
tensor of the second rank, by means of $B^{(q)}$ and its conjugate, which satisfies the continuity equation.
Such a tensor can then be regarded as the energy-momentum tensor of the field.
\marginnote{p.9}
First we need to specify more precisely what we mean by the conjugate tensor $C^*$ of a tensor $C$, which also reveals what 
is a "real" tensor.

Let the number of indices of $C$ equal to 4 be $n$ and $\bar{C}$ be the complex conjugate  of $C$, then 
\begin{equation}\label{eq2.1}
C^*=(-1)^n\bar{C}
\end{equation}
is the tensor conjugated to $C$. A tensor is "real" if $C^*=C$. In this sense, our coordinates $(X, Y, Z, ict) = (X_i)$ 
are "real" vectors. Likewise, the electromagnetic field $(\cH,\cF)$ is a "real" field. We now form the tensors
$T^{(q)}_{kl}$ by means of $B^{(q)}$ and $B^{(q-1)}$, as well as their conjugate:

\begin{alignat}{1}
T^{(1)}_{kl}&=\frac{\kappa^2}{2}(A^*_{ir\dots k}A_{ir\dots l}+A^*_{ir\dots l}A_{ir\dots k})
+\frac12(B^{(1)*}_{i\dots r[tk]}B^{(1)}_{ir\dots[tl]}+B^{(1)*}_{i\dots[tl]}B^{(1)}_{i\dots[tk]})\nonumber\\
&\quad-\frac12\delta_{kl}(A^*_{ir\dots}A_{ir\dots}\cdot\kappa^2 +\frac12B^{(1)*}_{i\dots[tr]}B^{(1)}_{i\dots[tr]}) \nonumber\\
\nonumber\\
T^{(q)}_{kl}&=\frac{\kappa^2}{2}(B^{*(q-1)}_{[rs]\dots tk}B^{(q-1)}_{[rs]\dots tl}+B^{*(q-1)}_{[rs]\dots tl}B^{(q-1)}_{[rs]\dots tk}) 
+\frac12(B^{(q)*}_{[rs]\dots [tk]\dots m}B^{(q)}_{[rs]\dots [tl]\dots m}\nonumber\\
&\quad +B^{(q)*}_{[rs]\dots [tl]\dots m}B^{(q)}_{[rs]\dots [tk]\dots m})-\frac12\delta_{kl}(\kappa^2B^{(q-1)*}_{[rs]\dots t}B^{(q-1)}_{[rs]\dots t}\nonumber\\
&\quad+\frac12B^{(q)*}_{[rs]\dots [tm]\dots n}B^{(q)}_{[rs]\dots [tm]\dots n}).\nonumber
\end{alignat}
These tensors satisfy, on the basis of the differential equations for the $B^{(q)}$, the continuity equation:
$$
\frac{\partial T_{kl}^{(q)}}{\partial x_k}=0.
$$
It is now necessary to require that the total energy of the field is positive, which means that the integral 
$\int T^{(q)}_{44}dv$ must be definite over the whole space, since this represents the total energy of the field except 
for the sign. This is indeed the case, as we shall first show for the tensor $T^{(1)}_{ik}$, which is composed of $A$
and $B^{(1)}$.For this purpose, we decompose $A_{ik\dots}$ into plane waves
\setcounter{equation}{2}
\begin{equation}\label{eq2.3}
A_{ik\dots r}(\vec{x},t)=\frac1{\sqrt{V}}\sum_k A_{ik\dots r}^+(k)e^{ikx+i\omega(k) t}+A_{ik\dots r}^-(k)e^{ikx-i\omega(k)t}.
\end{equation}
where
$$
\omega(k)=+ick_4,\qquad k_ik_i=-\kappa^2.
$$
\marginnote{p.10}

Further, the $A^+$ and $A^-$, as a result of the equation (\ref{eq1.3}), satisfy the equation 
\begin{equation}\label{eq2.4}
\sum_{i=1}^3k_i A_{i\dots}^+ +k_4 A_{4\dots}^+ =\sum_{i=1}^3 k_i A_{i\dots}^- - k_4 A_{4\dots}^-=0.
\end{equation}
If we consider this, express the $B^{(1)}$ by the $A^+$ and $A^-$ and insert this into the integral 
$\int T^{(1)}_{44}dV$, we get
\begin{equation}\label{eq2.5}
-\int T^{(1)}_{44}dV=\sum_k|k_4|^2\{A^{+*}_{ir\dots}(k)A^+_{ir\dots}(k)+A^{-*}_{ir\dots}(k)A^-_{ir\dots}(k)\}=E.
\end{equation}
The energy is shown here as the sum of the energies of the individual Fourier components. It is therefore sufficient
to show that the contribution of each individual component is positive definite. It is thus to show that $A^{+*}_{ik\dots}(k)A^+_{ik\dots}(k)$ is positive definite. 
Consider this expression in the rest frame of $k_i$, where $(k_i) = (0,0,0, i \kappa)$. 
Because of the equation (\ref{eq2.4}), all the components 
of $A^+_{ik\dots}$ where an index is equal to 4 vanishes. For the others, according to (\ref{eq2.1})
$$
A^*_{ik\dots}=\bar{A}_{ik\dots}
$$
So that in the rest frame the energy of a plane wave assumes the positive definite form 
$$
\bar{A}_{ik\dots}A_{ik\dots}.
$$

Since in another reference system the energy multiplies by the positive factor 
$$
\frac1{\sqrt{1-\beta^2}}
$$
it follows that the total energy of our field is positive definite. Also for the question discussed here, 
the existence of a rest frame is essential for each plane wave. If $\kappa$ is set to zero, the energy 
is never negative, as we shall see below, but can disappear without the field tensors $A_{ik\dots}$ being zero.

The more general tensor $T^{(q)}$ can be treated in the same way as in the tensor $T^{(1)}$. One then finds 
$$
-\int T^{(q)}_{44}dV=\sum_k|k_4|^2\{B^{+(q-1)*}_{[il]\dots m\dots}B^{+(q-1)}_{[il]\dots m\dots}+B^{-(q-1)*}_{[il]\dots m\dots}B^{-(q-1)}_{[il]\dots m\dots}\}
$$
or, if we express the $B^{(q-1)}$ by the $A$:
$$
-\int T^{(q)}_{44}dV=(-2\kappa^2)^{q-1}\sum_k|k_4|^2\{A^{*+}_{il\dots}A^+_{il\dots}+A^{-*}_{il\dots}A^-_{il\dots}\}.
$$

\marginnote{p.11}

All energy-momentum tensors therefore give rise to the same total energy, except for the factor $(-2 \kappa^2)^{q-1}$.
However, the localization of the energy in the field is largely unclear, depending on the chosen $(q)$. Moreover, 
$T^{(q)}_{44}$ is not positive definite if $f> 1$. However, these properties of the fields considered here do not
seem to be sufficient to exclude particles with spin$> 1$. In the force-free cases the energy tensor is not 
determined unambiguously; instead, there are linearly independent ways to localize the energy, but the value of the total 
energy remains unchanged. To what extent this ambiguity can be restricted by the introduction of interactions with other 
fields, or whether it leads to more serious physical difficulties in this case, a separate investigation must be reserved.

In addition to the energy-momentum tensor $T^{(q)}_{kl}$, a vector can also be formed from $B^{(q)}$ and
$B^{(q-1)}$, which also satisfies the continuity equation and can be regarded as the current vector of the field: 
$$
S^{(q)}_v=
\frac1{2i}\{B^{*(q-1)}_{[ik]\dots ml\dots}B^{(q)}_{[ik]\dots[mv]l\dots}
-B^{(q-1)}_{[ik]\dots ml\dots}B^{(q)*}_{[ik]\dots[mv]l\dots}\}.
$$
This vector is also "real." If the field tensors $B^{(q)}$ are "real" quantities, the current vectors disappear identically.
Given a field, the current and charge density can be defined again in $f$ different ways, which are equivalent
in the  force-free case. The total charge $\int s^{(q)}_4dV$ is again in all cases, up to a factor, the same: 
$$
e=\sum_k|k_4|\{A^{+*}_{il\dots}(k)A^+_{il\dots}(k)-A^{-*}_{il\dots}(k)A^-_{il\dots}(k)\}.
$$
However, whether the so-defined current and charge density can be interpreted as electrical current and charge density,
can not be decided here since this depends on the manner in which the electric field is introduced. 
Since $\vec{p}$ is replaced by $\vec{p}-\frac ec\vec{\phi}$ in all equations, this is not possible in
the general case, since the algebraic relations which the $A_{ik\dots}$ have to fulfill if $f> 1$ are then 
in difficulty. Only for $f = 1$, where the spin condition (\ref{eq1.2}) falls, we arrive at a contradiction-free
theory which is identical with that discussed by Proca.

\marginnote{p.12}
\subsection{Relation of the theory developed here to Dirac's relativistic equations}
Dirac has given differential equations which have the following form in spinor spelling:
\begin{alignat}{1}
p^{\dr\kappa}a^{\dt\dn\dots}_{\kappa\lambda\mu\dots}=\kappa b^{\dr\dt\dn\dots}_{\lambda\mu\dots}\label{eq3.1}\\
\nonumber\\
p_{\dn\rho}b^{\dn\dt\dm\dots}_{\lambda\gamma\dots}=\kappa a^{\dt\dm\dots}_{\rho\lambda\gamma\dots}\label{eq3.2}
\end{alignat}
Here
\begin{itemize}
\item[] $a$ has $2k$ undotted indices and $2l-1$ dotted ones.
\item[] $b$ has $2l$ dotted indices and $2k-1$ undotted ones.
\item[] $k$ and $l$ are integer or half-integer numbers.
\end{itemize}
$a$ and $b$ are symmetric in dotted and undotted indices, and therefore irreducible in Lorentz transformations, 
where they are transformed according to the representation $\theta_{k,l-\frac12}$ and $\theta_{k-\frac12,l}$, 
respectively. If $k + l-\tfrac12 = f$ is integer, then these equations are equivalent to the tensor equations we have looked at.
All equations with a given $f$ describe the same wave field, the division of $f + \frac12$ into two summands corresponding
to the various possibilities characterized by the index $(q)$ in the tensor field.

If, in particular for $a^{\dl\dm}_{\delta\rho}$, a spinor, which has the same number of dotted as undotted indices, 
that is, $k = l-\frac12$, we see that this quantity $a$ is identical with the tensor $A_{ik\dots}$. 
One can use the Pauli matrices $\sigma^k_{\dr\rho}$ to transform an undotted and a dotted 
index $\dr$, $\rho$ into a vector index $k$.
The resulting tensor is then symmetric in all the indices and its traces disappear, since the trace 
$\epsilon^{\mu\nu}a_{\mu\nu} = a^\mu_{\ \mu}$ of a symmetric spinor $a_{\mu\nu} = a_{\nu\mu}$ disappears.
Further, it also satisfies the differential equations (\ref{eq1.1}) and (\ref{eq1.3}). (\ref{eq1.3}) follows from 
the spinor equation 
\begin{equation}\label{eq3.3}
p^{\dr\kappa}a^{\dt\dots}_{\dr\kappa\lambda\mu\dots}=0
\end{equation}
 which results from contraction from (\ref{eq3.1}); It must be noted again that the trace of the symmetric spinor 
 $b^{\dl\dm\dots}_{\delta\rho\dots}$ vanishes.

From $a^{\dr\dm\dots}_{\lambda\kappa\dots}$ we now form new spinors $b^{(q)}$ according to 
\begin{alignat*}{1}
\kappa b^{(1)\dr\dt\dm\dots}_{\lambda\dots} &= p^{\dr\kappa} a^{\dt\dm\dots}_{\kappa\lambda\dots}\\
\\
\kappa b^{(-1)\dn\dots}_{\mu\kappa\lambda\dots} &= p_{\mu\dr} a^{\dr\dn\dots}_{\kappa\lambda\dots}
\end{alignat*}

\marginnote{p.13}
and in general
\begin{alignat}{1}
\kappa b^{(q+1)\dr\dm\dots}_{\lambda\dots} &= p^{\dr\kappa} b^{(q)\dm\dots}_{\kappa\lambda\dots}\nonumber\\
\nonumber\\
\kappa b^{(q-1)\dt\dots}_{\mu\kappa\lambda\dots} &= p_{\mu\dr} b^{(q)\dr\dt\dots}_{\kappa\lambda\dots}\label{eq3.4}
\end{alignat}
$(q)$ is in each case equal to half the difference between the dotted and undotted indices and therefore runs from $-f$ 
to $+f$. $a^{\dl\dm\dots}_{\delta\rho\dots}$ has $f$ dotted and $f$ undotted indices and transforms according to the
irreducible representation of the Lorentz group $\theta_{\frac f2,\frac f2}$. $b^{(q)\dl\dm\dots}_{\delta\rho\dots}$ has 
$f + q$ dotted and $f-q$ undotted indices and transforms itself in the case of actual Lorentz transformations 
after the irreducible representation $\theta_{\frac{f-q}{2},\frac{f+q}{2}}$. 
For reflections $b^{(q)}$ exchanges with $b^{(-q)}$. Thus the $b^{(q)}$, in contrast to the $B^{(q)}$, are irreducible
quantities. They correspond to tensors with $(q)$ index pairs in which the tensor is skew and self-dual according to 
\begin{alignat*}{1}
\epsilon^{iklm}F^{(q)}_{\dots[lm]}&=F^{(q)}_{[ik]\dots}\qquad\text{if }q>0\\
\\
-\epsilon^{iklm}F^{(q)}_{\dots[lm]\dots}&=F^{(q)}_{\dots[ik]\dots}\qquad\text{if }q<0.
\end{alignat*}
In the case of reflection, we exchange for $F^{(q)}$ with $F^{(-q)}$, and we can form $f$ energy-momentum tensors, 
just as with the $B^{(q)}$, which are symmetrical and satisfy the continuity equation
\begin{alignat*}{1}
t^{(q-1)}_{\dr\delta,\beta\dn}
&=
b^{(q)*}_{\dr\beta}b^{(-q)}_{\dn\delta}+b^{(-q)*}_{\dr\beta}b^{(q)}_{\dn\delta}+b^{(q)*}_{\dn\delta}b^{(-q)}_{\dr\beta}+b^{(-q)*}_{\dn\delta}b^{(q)}_{\dr\beta}\\
&+b^{(q-1)*}_{\dn\dr}b^{(-q-1)}_{\beta\delta}+b^{(-q-1)*}_{\dr\dn}b^{(q-1)}_{\beta\delta}
+b^{(q+1)*}_{\beta\delta}b^{(-q+1)}_{\dr\dn}+b^{(-q+1)*}_{\beta\delta}b^{(q+1)}_{\dr\dn}.
\end{alignat*}
The non-registered indices are to be contracted according to: 
$$
a^*a=a^{*\dr\dn\dots}_{\alpha\delta\dots}a^{\alpha\delta\dots}_{\dr\dn\dots}.
$$
On the basis of the equations (\ref{eq3.4}) for $t^{(q-1)}$: 
$$
p^{\dr\delta}t^{(q-1)}_{\dr\delta,\dn\gamma}=0.
$$

Since the tensor formulation and the spinor notation are mathematically equivalent in this case, 
it follows that all the tensors we have considered above are linear combinations of corresponding spinors.
We can therefore forgo a further discussion of the equations (\ref{eq3.4}) and refer to the theory of tensors.

\marginnote{p.14}
\subsection{Quantization of the field theory}
In order to be able to assign particles to the field theories discussed above, we have to establish the Lorentzian
variant of commutation relations between the field magnitudes. It is sufficient to do this for the $A_{ik\dots}$,
since those for the other magnitudes then follow by differentiation. Since these are theories of the case free of forces,
it is expedient to establish the commutation relations in four-dimensional form, analogous to that given by Jordan and Pauli
in the charge-free electromagnetic field. One has the advantage, then, that the Lorentz variance is secured from the start.
The commutation relations should have the effect that the energy of a plane wave becomes an integer multiple of 
$|k_4|=\sqrt{k^2+m^2}$, and the equation 
\begin{equation}\label{eq4.1}
\dot{f}=\frac1i[\overline{H},f]
\end{equation}
should also hold for all quantities which do not contain the time explicitly.
 
In the case of $f = 1$, the task thus set can be solved by introducing $2f + 1 = 3$ independent amplitudes to each wave
vector $k_i$  by introducing longitudinal and transverse waves, and only requires these 
commutation relations. At higher spins, however, a corresponding method does not appear to us to be applicable 
without completely destroying the symmetry of the problem, which, on the one hand, makes the assessment of the 
invariance of the resulting commutation relations impossible. But, instead of such a procedure, it is possible to establish
commutation relations between the $A_{ik\dots}$ and the $A^*_{ik\dots}$, of which all the equations to which the 
$A_{ik\dots}$ suffice are fulfilled identically. This automatically takes account of the secondary conditions 
(\ref{eq1.2}) and (\ref{eq1.3}), thus obtaining the correct number of independent commutation relations.

We therefore consider the following as the commutation relations for the symmetric tensor $A(i_1\dots i_f)$:
\begin{alignat}{1}
\frac1i[&A(i_1\dots i_f),A^*(i_1'\dots i_f')] 
= K\{\Sigma P(i'_k)R(i_1i'_1)\dots R(i_fi'_f)\label{eq4.2}\\
&-\frac1{2+\frac{f(f-1)}{2}}
\sum_{l>m}^f R(i_li_m)\Sigma P(i'_k)R(i_1i'_1)\dots R(i'_li'_m)\dots R(i_fi'_f)\}D(x).\nonumber
\end{alignat}
Here, $A(i_k)$ is to be taken at the position $\xi+\frac x2$, $A^*(i'_k)$ at the position $\xi-\frac x2$.
\marginnote{p.15}
It means: $\Sigma P(i_k')$ the sum over all permutations of the indices $i'_k$. $R(i_ki_l)$ is the operator

\begin{equation}
R(i_ki_l)=\delta_{i_ki_l}-\frac1{\kappa^2}\frac{\partial^2}{\partial x_{i_k}\partial x_{i_l}}\label{eq4.3}
\end{equation}
which is always to be applied to the invariant $D$-function of Jordan and Pauli, which is defined as follows: 
\begin{equation}\label{eq4.3b}
D(x)=D(\bx,t)=\frac12\sum_k\frac1V
\left\{\frac{e^{i\bk\bx+i\omega(k)t}}{i\omega(k)}-\frac{e^{i\bk\bx-i\omega(k)t}}{i\omega(k)}\right\} \tag{4.3b}
\end{equation}
D (x) satisfies the wave equation 
$$
\square D=\kappa^2 D
$$
further
$$
D(\bx,0)=0,\quad\left(\frac{\partial D}{\partial t}\right)_{t=0}=\delta(\bx).
$$

Since $R(\alpha\beta)$ is always applied to $D(x)$, which satisfies the wave equation, the following relations hold:
\begin{alignat}{1}\label{eq4.4}
\frac{\partial R(\alpha\beta)}{\partial x_\alpha}=R(\alpha\beta)\frac{\partial}{\partial x_\alpha}=0,
&\qquad
R(\alpha\beta)R(\beta\gamma)=R(\alpha\gamma)\nonumber\\
R(\alpha\alpha)&=3
\end{alignat}
The constant factor $K$ in (\ref{eq4.2}) has to be chosen so that the eigenvalues of the charge become integers.

The commutation relations (\ref{eq4.2}) now fulfill the equations (\ref{eq1.1}) to (\ref{eq1.3}), and they are symmetric
in the primed and unprimed indices, since all permutations of the $i_k'$ are summed. 
Their divergence is zero because of (\ref{eq4.4}). To see that the traces also vanish, consider two terms from the
sum $\sum_{l>m}$ in (\ref{eq4.2}):
\begin{alignat*}{1}
\alpha)\quad& R(i_ri_k)\Sigma P(i'_n)R(i_1i'_1)\dots R(i'_ri'_k)\dots R(i_fi'_f)\\
\beta)\quad & R(i_ki_t)\Sigma P(i'_n)R(i_1i'_1)\dots R(i'_ki'_t)\dots R(i_fi'_f)
\end{alignat*}
Now form the trace over $(i_ki_r)$ and observe the equations (\ref{eq4.4}).
We thus obtain from 
\begin{alignat*}{1}
\alpha)\quad& 3\Sigma P(i'_n)R(i_1i'_1)\dots R(i'_ki'_r)\dots R(i_fi'_f)\\
\beta)\quad & \Sigma P(i'_n)R(i_1i'_1)\dots R(i_ti'_r)\dots R(i'_ki'_t) \dots R(i_fi'_f)
\end{alignat*}
In both expressions, all $i'_n$ occur; however, $i_k$ and $i_r$ are missing.

\marginnote{p.16}
We can now take the $R(i_ni'_l)$ with respect to the remaining $i_n$ in both expressions. Since all the permutations 
of the $i'_n$ occur, $\alpha$ gives up to the factor 3 the same as $\beta$. If we now form the trace over ($i_ki_r$) in 
$$
\gamma)\quad \Sigma P(i'_k)R(i_1i'_1)\dots R(i_fi'_f),
$$
we obtain the same result as in $\beta)$. In the sum $\sum_{l>m}$ we have $\frac{f(f-1)}{2}$ summands, of which
the triple of the others is the result of forming the trace. Therefore, in order to satisfy equation (\ref{eq1.2}), 
the factor 
$$
\frac{1}{\frac{f(f-1)}{2}+2}
$$
must be placed before the sum $\sum_{l>m}$.

We now have to show that the equation $\frac1i\dot{f}=[\overline{H}, f]$ holds, and that the energy has the 
proper eigenvalues. For this purpose we decompose the $A_{ik\dots}$ into plane waves according to (\ref{eq2.3}).
It follows from the commutation relations that $A(k)$ commutes with $A(k')$ if $k_i\not= k'_i$. $A^+$ also commutes
with $A^-$. For given $k_i$ there are now $2f + 1$ linearly independent linear combinations of the $A^+_{ik\dots l}(k)$, 
which we call $A_\omega$. The commutation relations of $A_\omega$ may be as follows:
\begin{equation}\label{eq4.5}
[A_\omega,A^*_{\omega'}]=f_{\omega\omega'}.
\end{equation}
We now consider the $A_\omega$ again in the rest frame belonging to $k_i$. There the $A_\omega$, which transform as 
the $A^*_\omega$ after the irreducible representation $\theta_f$ of the rotation group, span also a space $\cR_{2f+1}$.
Since the right-hand side of the commutation relations (\ref{eq4.2}) satisfies the same relations as $A_\omega$,
the matrix $f_{\omega\omega'}$ has the property that it maps the vectors $X_\omega$ in $\cR_{2f+1}$ to vectors 
$Y_{\omega'}=X_\omega f_{\omega\omega'}$ which again span an irreducible space which, since the $Y_{\omega'}$ 
are not all null, has again dimension $2f+1$.
Therefore, $f_{\omega\omega'}$ has $2f+1$ eigenvalues different from zero.
Now we choose the $A^*_\omega{}'$ so that the representation rendered by it becomes unitary.

Since the commutation relations are invariant by rotations, they have the form, if the $A_\omega$ are suitably normalized
\begin{equation}\label{eq4.6}
[A_\omega,A^*_{\omega'}]=\delta_{\omega\omega'}.
\end{equation}
\vskip1pc
\marginnote{p.17}
The energy is also rotationally invariant and therefore is written into these $A_\omega$  invariant unit form
\begin{equation}\label{eq4.7}
E(k_i)=\sum_{\omega=1}^{2f+1}|k_4|CA^*_\omega A_\omega.
\end{equation}
The constant $K$ occurring in the commutation relations (\ref{eq4.2}) must now be determined so that $C=1$.

Now, according to (\ref{eq4.7})
$$
|k_4|A^{+*}_{ik\dots}A^+_{ik\dots}=\sum_{\omega}A^*_\omega A_\omega
$$
if $C=1$. It must therefore also apply 
$$
|k_4|[A^{+}_{ik\dots},A^{+*}_{ik\dots}]=\sum_{\omega}[A_\omega, A^*_\omega]=2f+1.
$$
From this, by comparison with (\ref{eq4.2})
\begin{alignat*}{1}
\sum_{i,k}\delta_{i_k,i'_k}&\left\{\Sigma P(i'_k)R(i_1i'_1)\dots R(i_fi'_f)\phantom{\sum_l}\right.\\
&\left. -\frac1{2+\frac{f(f+1)}{2}}\sum_{l>m}R(i_li_m)\Sigma P(i'_k)R(i_1i'_1)\dots
\right\}D(x)=\frac{2f+1}{K}D(x)
\end{alignat*}
the constant $K$ is determined.

The commutation relation now has the property that it follows from them that the eigenvalues of the energy
of a plane wave are integral multiples of $k_4$; Because for $A^-$, the corresponding one follows exactly
the same way. These can be represented by $2f+1$ variables $B_\omega$, which satisfy the commutation relations 
\begin{equation}\label{eq4.8}
[B_\omega,B^*_{\omega'}]=-\delta_{\omega\omega'}
\end{equation}
the minus sign resulting from the minus sign in the definition of the $D$ function. Since every quantity $f$ of 
$A_\omega e^{i\omega t}$ and $B_\omega e^{-i\omega t}$, which does not explicitly contain time, can be constructed linearly, 
the equation 
$$
\frac1i\dot{f}=[\overline{H},f]
$$
follows for every $f$, since it applies to $A_\omega$ and $B_\omega$.

Thus it is shown that the commutation relations (4.2) form the solution of the initial task.

The spin-form commutation relations are found in the second part, which deals with the case of half-integer spins, 
where we compare them with those for this case.

\marginnote{p.18}
\section{Half-integer spin}
\setcounter{subsection}{4}
\subsection{Field quantities and wave equations}

The fields to which particles with a half-numbered spin $f$ ($f=\frac13,\frac32,\frac52\dots$) can be assigned can,
of course, not be represented by tensors, but by spinors ("half-tensors"). 
In spinor formulations, the theories of integer and half-integer spin are very similar in many respects, 
although characteristic differences occur, which in particular concern the construction of the energy 
tensors and the commutation relations. Otherwise, however, most of the conclusions which we made in the case of 
an integer case can also be applied to the following theory, so that we can reduce ourselves here.

For the particles described by a spinor-field to belong to a half-numbered spin, the spinors must have an odd number of indices, while in the integer case the index number is even.

We therefore proceed from a spinor $a^{\dl\dm\dots}_{\delta\rho\dots}$ which has $2k$ undotted and $2l-1$ dotted indices, 
where $2k + 2l-1 = 2f$ is an odd number. $k$ and $l$ are integer or half-integer numbers.

$a^{\dl\dm\dots}_{\delta\rho\dots}$ is to satisfy the second-order wave equation: 
\begin{equation}\label{eq5.1}
\square a^{\dl\dm\dots}_{\delta\rho\dots}=\kappa^2 a^{\dl\dm\dots}_{\delta\rho\dots}
\end{equation}
where $\kappa$ is again a characteristic wave number which determines the mass of the particles according to 
$\kappa=\frac{mc}{\hbar}$.
Furthermore, $a^{\dl\dm\dots}_{\delta\rho\dots}$ should be symmetric in dotted and undotted indices, which can 
also be formulated as follows: Let all traces of $a^{\dl\dm\dots}_{\delta\rho\dots}$ vanish: 
\begin{equation}\label{eq5.2}
\epsilon_{\dn\dl}a^{\dl\dn\dots}_{\delta\rho\dots}=0,\qquad \epsilon^{\delta\rho}a^{\dl\dm\dots}_{\delta\rho\dots}=0
\end{equation}
where $\epsilon_{\dl\dm}$ is the oblique matrix $\left(\begin{smallmatrix}0 & 1\\-1 & 0\end{smallmatrix}\right)$.
Further, let a satisfy the constraints 
\begin{equation}\label{eq5.3}
\epsilon_{\dn\dm}p^{\dn\rho}a^{\dl\dm\dots}_{\rho\delta\dots}=0
\end{equation}
These mean that the spinors $p^{\dn\rho}a^{\dl\dm\dots}_{\rho\delta\dots}$ and $p_{\dn\tau}a^{\dn\dm\dots}_{\rho\delta\dots}$ 
are symmetric in dotted and undotted indices.

Since $a^{\dl\dm\dots}_{\gamma\delta\dots}$ is given, if one knows how many dotted and how many undotted indices are equal 
to one, then it has $(2k + 1) 2l$ linearly independent components, which transform themselves in Lorentz transformations 
after the irreducible representation $\theta_{k,l-\frac12}$. Now it remains to be shown again that,
\marginnote{p.19}
due to the constraint (\ref{eq5.3}), $2f+1$ independent plane waves exist for each wave vector.
For this purpose, we return to the rest frame of $k_i$. There, $p^{\dn\rho}$ has the form $k_4\delta^{\dn\rho}$, 
which transforms as $k_4\delta^\rho_\gamma$ under spatial rotations. 
Likewise, $a^{\dl\dm\dots}_{\rho\delta\dots}$ rotates with rotations like the spinor
$a'_{\rho\delta\dots\lambda\mu\dots}$\footnote{See also van der Waerden\cite{waerden}, 
{\em Di gruppentheoret. Methode}, p. 81.}.
Equation (\ref{eq5.3}) is then 
\begin{equation}\label{5.3a}
\epsilon^{\gamma\lambda}a'_{\gamma\delta\dots\lambda\mu\dots}=0\tag{5.3a}
\end{equation}
and states that the spinor $a'$ is symmetric in the indices $\lambda$ and $\mu$, that is, in all indices.
In the rest frame, therefore, $a^{\dl\dm\dots}_{\rho\delta\dots}$ is equivalent to a symmetric spinor
 $a'_{\gamma\delta\dots\lambda\mu\dots}$ of rank $2k+2l-1=2f$, and therefore has $2k+2l=2f+1$ linearly 
independent components which transform with one another in the case of rotations under the irreducible 
representation $\theta_f$ of the rotation group.
This shows that the spin to be assigned to the wave field is $f$. If $\kappa\not=0$, there is no rest frame,
and these in turn become invalid.

The differential equations (\ref{eq5.1}) and (\ref{eq5.3}) can now be replaced by a first-order equation system:
\begin{alignat}{1}
p^{\dn\rho}a^{\dl\dm\dots}_{\rho\delta\dots}=\kappa b^{\dn\dl\dm\dots}_{\delta\dots}\nonumber\\
\label{eq5.4}\\
p_{\dn\rho}b^{\dn\dl\dm\dots}_{\delta\dots}=\kappa a^{\dl\dm\dots}_{\rho\delta\dots}\nonumber
\end{alignat}
(\ref{eq5.3}) implies that $b^{\dn\dl\dm\dots}_{\delta\dots}$ is again a symmetric spinor satisfying the second 
order wave equation and the secondary condition (\ref{eq5.3}).
Hence all equations with fixed $k+l$ can be derived by differentiation; they all describe the same wave field.
Let $a^{(0)\dl\dm\dots}_{\rho\delta\dots}$ be the spinor $a$ for which $k=l$, which thus has $2k$ undotted and 
$2k-1$ dotted indices.
It satisfies the equations
\begin{alignat}{1}
p^{\dn\rho}a^{(0)\dl\dm\dots}_{\rho\delta\dots}=\kappa b^{(0)\dn\dl\dm\dots}_{\delta\dots}\nonumber\\
\label{eq5.5}\\
p_{\dn\rho}b^{(0)\dn\dl\dm\dots}_{\delta\dots}=\kappa a^{(0)\dl\dm\dots}_{\rho\delta\dots}\nonumber
\end{alignat}
$b^{(0)}$ has $2k$ dotted and $2k-1$ undotted indices.
From $a^{(0)}$ we now form the spinor $a^{(1)}$
$$
p_{\dl\tau}a^{(0)\dl\dm\dots}_{\rho\delta\dots}=\kappa a^{(1)\dm\dots}_{\tau\rho\delta\dots}
$$
And similarly from $a^{(1)}$ the spinor $a^{(2)}$. These spinors are again symmetric because of the equation (\ref{eq5.3}).
\marginnote{p.20}
In General,
\begin{alignat}{1}
p_{\dl\tau}a^{(q)\dl\dm\dots}_{\rho\delta\dots}=\kappa a^{(q+1)\dm\dots}_{\tau\rho\delta\dots}\nonumber\\
\label{eq5.6}\\
p^{\dl\tau}a^{(q+1)\dm\dots}_{\tau\rho\delta\dots}=\kappa a^{(q)\dl\dm\dots}_{\rho\delta\dots}\nonumber
\end{alignat}
We thus obtain the sequence of symmetric spinors $a^{(0)}, a^{(1)}\dots,a^{(f-\frac12)}$ which all satisfy 
the wave equation, and, up to $a^{(f-\frac12)}$, also the subcondition (\ref{eq5.3}), $a^{(q)}$ has $f+q+\frac12$ 
undotted and $f-q-\frac12$ dotted indices. 
We also form the spinors $b^{(1)}\dots b^{(f-\frac12)}$ according to
\begin{alignat}{1}
p^{\dl\tau}b^{(q)\dm\dn\dots}_{\tau\rho\dots}=\kappa b^{(q+1)\dl\dm\dn\dots}_{\rho\dots}\nonumber\\
\label{eq5.7}\\
p_{\dl\tau}b^{(q+1)\dl\dm\dn\dots}_{\rho\dots}=\kappa b^{(q)\dm\dn\dots}_{\tau\rho\dots}\nonumber
\end{alignat}
where $b^{(q)}$ has $f+q+\frac12$ dotted and $f-q-\frac12$ undotted indices.
For reflections, $b^{(q)}$ is interchanged with $a^{(q)}$ such that (\ref{eq5.6}) and (\ref{eq5.7}) are mutually
mirror-invariant.
The system of equations (\ref{eq5.5}), \ie\ the case $k=l$, is, therefore, excellent inasmuch as this system of
equations is itself inversely mirror-invariant.
In contrast to the case of integer spins, there is no spinor $a$, which would be invariant to reflection.

\subsection{Energy-momentum tensor and current vector}

Again, by means of the $a^{(q)}$ and $b^{(q)}$ tensors, we can form a second rank tensors and vectors, which can 
be interpreted as the energy-momentum tensors and current vectors. In this case, however, it is found that, as 
is known from case $f=\frac12$, the energy is not positive definite, but this is the case for the total charge
because of the secondary condition (\ref{eq5.3}). We first form the vector $s^{(q)}_{\dl\beta}$
\begin{alignat}{1}
s^{(0)}_{\dl\beta}&=a^{(0)*\rho\delta\dots}_{\dl\dn\dt}a^{(0)\dn\dt\dots}_{\beta\rho\delta\dots}+b^{(0)*\rho\delta\dots}_{\beta\dn\dt\dots}b^{(0)\dn\dt\dots}_{\dl\rho\delta\dots}\nonumber\\
\label{eq6.1}\\
s^{(q)}_{\dl\beta}&=\frac12(a^{(q)*}_{\dl}b^{(q-1)}_{\beta}+a^{(q)}_{\beta}b^{(q-1)*}_{\dl}
+a^{(q-1)}_{\dl}b^{(q)*}_{\beta}+a^{(q-1)*}_{\beta}b^{(q)}_{\dl})\nonumber
\end{alignat}
(The omitted indices are contracted as in $s^{(0)}_{\dl\beta}$). $s^{(q)}_{\dl\beta}$ satisfies the continuity equation
$$
p^{\dl\beta}s^{(q)}_{\dl\beta}=0,
$$
as can be easily recalculated, if we consider that $a_\lambda b^\lambda=-a^\lambda b_\lambda$.
There are, therefore, different equivalent ways of defining the current, in the absence of a force. 
Let us again show that all these possibilities lead to the same value of the total charge 
$\int (s^{(q)}_{1\dot{1}}+s^{(q)}_{2\dot{2}})dV$, which in this case is definite.

\marginnote{p.21}
First we consider $\rho^{(0)}_{\dl\beta}$: Since $\rho_{\dl\beta}$ satisfies the continuity equation,
the total charge is constant in time and the integral is divided into the sum over the charge contributions 
of the individual plane waves with a given wave number and frequency.
We therefore consider the charge density of a plane wave. If this is positive, it follows that the total charge is also positive. We consider the plane wave in its rest frame. There the secondary condition (\ref{eq5.3}) states that 
$$
a^{\dot{1}\dn\dots}_{2\alpha\dots}=a^{\dot{2}\dn\dots}_{1\alpha\dots},\quad b^{\dot{1}\dt\dots}_{2\beta\dots}=b^{\dot{2}\dt\dots}_{1\beta\dots}.
$$
Therefore the following equation applies in the rest frame:
\begin{equation}\label{eq6.2}
a^{(0)*\rho\dots}_{\dot{1}\dm\dots}a^{(0)\dm\dots}_{1\rho\dots}=a^{(0)*\rho\dots}_{\dot{1}\dm\dots}a^{(0)\dr\dots}_{1\mu\dots}
\end{equation}
$a^{(0)*\rho\dots}_{\dot{1}\dn\dots}$ is the complex conjugate of $a^{(0)\dr\dots}_{1\nu\dots}$, so the right-hand side 
of (\ref{eq6.2}) has the positive definite form $\sum a^*a$. The charge density of a plane wave consists of summands 
of the type (\ref{eq6.2}), so the total charge is positive definite.

To discuss the more general case 
$$
\int(s^{(q)}_{1\dot{1}}+s^{(q)}_{2\dot{2}})dV
$$
it suffices to consider a particular plane wave of given wave number and frequency in its rest frame. 
Let us assume that the temporal dependence of the amplitudes $a^{(q)}, b^{(q)}$ is $e^{i\omega t}$; Then that of 
$a^{(q)*},b^{(q)*}$ is equal to $e^{-i\omega t}$. $a^{(q)}$ and $b^{(q-1)}$ are now obtained from $a^{(0)}$ by applying 
the operator $\frac1\kappa p^{\dl\beta}$  $q$ times.
But this is equal to $i\delta^{\dl\beta}$ in the rest frame for $a^{(0)}$ and $b^{(0)}$, and $-i\delta^{\dl\beta}$ for $a^{(0)*}$ and $b^{(0)*}$. From this it immediately follows that the equation holds: 
$$
\int(s^{(q)}_{1\dot{1}}+s^{(q)}_{2\dot{2}})dV=\int(s^{(0)}_{1\dot{1}}+s^{(0)}_{2\dot{2}})dV.
$$
This shows that all the vectors $s^{(q)}_{\alpha\dm}$ lead to the same total charge. 
However, the localization of the charge depends again on $(q)$.

We can also form spinors which correspond to tensors of the second rank, and satisfy the continuity equation.
These can again be regarded as energy-momentum tensors of our field. We look at the spinor
\begin{alignat}{1}
t^{(0)}_{\dl\beta,\gamma\dt}&=\frac12
(a^{(0)*}_{\dl}p_{\dn\delta}a^{(0)}_\beta-a^{(0)}_\beta p_{\gamma\dt}a^{(0)*}_{\dl}
+b^{(0)*}_\beta p_{\gamma\dt}b^{(0)}_{\dl}-b^{(0)}_{\dl}p_{\gamma\dt}b^{(0)*}_\beta)\nonumber\\
t^{(q)}_{\dl\beta,\dt\gamma}&=\frac14\big(a^{(q)*}_{\dl}p_{\gamma\dt}b^{(q-1)}_\beta-b^{(q-1)}_\beta p_{\gamma\dt}a^{(q)*}_{\dl}
+b^{(q-1)*}_{\dl} p_{\dt\gamma}a^{(q)}_{\beta}-a^{(q)}_{\beta}p_{\dt\gamma}b^{(q-1)*}_{\dl} \label{eq6.3}\\
&\qquad +b^{(q)*}_{\beta}p_{\dt\gamma}a^{(q-1)}_\lambda-a^{(q-1)}_{\dl} p_{\dt\gamma}b^{(q)*}_{\beta}
+a^{(q-1)*}_\beta p_{\dt\gamma}b^{(q)}_{\dl}-b^{(q)}_{\dl}p_{\gamma\dt}a^{(q-1)*}_\beta\big)\nonumber
\end{alignat}

\vskip1pc
\marginnote{p.22}
$t^{(q)}_{\dl\beta,\dt\gamma}$ satisfies the equations 
$$
p^{\dl\beta}t^{(q)}_{\dl\beta,\dt\gamma}=0,\quad p^{\dt\gamma}t^{(q)}_{\dl\beta,\dt\gamma}=0.
$$
We can therefore form the spinor $\Theta{(q)}$ from $t^{(q)}$, according to 
$$
\Theta^{(q)}_{\dl\beta,\dt\gamma}=\frac12(t^{(q)}_{\dl\beta,\dt\gamma}+t^{(q)}_{\dt\gamma,\dl\beta}).
$$
The tensor $T_{kl}$ associated with $\Theta^{(q)}_{\dl\beta,\dt\gamma}$ 
$$
T_{kl}=\Theta^{(q)}_{\dl\beta,\dt\gamma}\sigma^{\dl\beta}_k\sigma^{\dt\gamma}_l
$$
is symmetric in $kl$ and satisfies the continuity equation 
$$
\frac{\partial T_{kl}}{\partial x_k}=0.
$$
$T_{kl}$ can therefore be regarded as the energy-momentum tensor of the field.
It is easy to see with the help of the Fourier decomposition of the field magnitudes, and again using the existence 
of a rest frame, that all $\Theta^{(q)}$ lead to the same value of the total energy, and that the contribution to 
the energy of plane waves varying with $e^{+i\omega t}$ has the opposite sign as the  contribution of waves that 
vary according to $e^{-i\omega t}$.
There are states of positive and negative energy, as is known from the Dirac theory of the electron.
We will therefore have to postulate that the particles assigned to the wave field fulfill the Paul exclusion 
principle, so that the energy can be made positive by analogy to the hole theory of Dirac.

\subsection{The matrices $u_\nu(k)$, $v_\nu(k)$ }

The spinors considered so far were all symmetrical. They are therefore given when the number of undotted and dotted 
indices equal to one is known.
We can therefore replace a spinor $a^{\dl\dm\dots}_{\delta\rho\dots}$ with $2k$ undotted and $2l$ dotted indices, 
which is symmetrical in these, by a quantity $A^{\dot{r}}_s$ whose indices $\dot{r}, s$ indicate how many of the dotted and 
undotted indices are one. $s$ then has $2k+1$ values, and $r$ has $2l+1$ values.
In order to obtain an assignment between the quantity $A^r_s$ and the symmetric spinor $a^{\dl\dm\dots}_{\rho\delta\dots}$, 
we have to search for a matrix which transfers from the spinor indices to the indices $r, s$. 
It is, of course, possible to assign a magnitude $A^{\dot{r}}_s$ to a spinor with no special symmetry properties, 
but the image generated in this way is not 
\marginnote{p.23}
unique, since $A^{\dot{r}}_s$ determines only the symmetrical part of the associated spinor.
Since a spinor without symmetry properties is transformed according to a product representation of the Lorentz group, 
which contains $2k$ times the factor $\theta_{\frac12,0}$ and $2l$ times the factor $\theta_{0,\frac12}$, 
but $A^{\dot{r}}_r$ is irreducible, the assignment of such a quantity to a non-symmetric spinor means 
the extraction of an irreducible subspace, namely that of the greatest number of dimensions from the representation 
belonging to the spinor.
We can therefore find a mapping which leads from spinor indices to an index $s$ by the reduction theory of 
representations.
We therefore consider the product representation of the rotation group $\theta_{\frac12}\times\theta_k$. $\theta_{\frac12}$ 
and $\theta_k$ are characterized by their infinitesimal transformations $\sigma^l$ and $\alpha^l(k)$.
The product presentation can be characterized by the matrix 
$$
A(k)=\sum_{l=1}^3\sigma^l\times\alpha^l(k)
$$
already considered by Dirac.
The matrix $A(k)$ is interchangeable with all matrices of the product representation. If they are applied to
the principal axes, the representation $\theta_{\frac12}\times\theta_k$ decomposes into irreducible constituents.
The reduction $\theta_{\frac12}\times\theta_k$ is thus reduced to the principal axis transformation of 
$A(k)$\footnote{See also: H. Casimir and B. L. van der Waerden, Math. Ann. 111 (1935), p.1}. A matrix which maps $A(k)$ on major axes has already been given by Dirac. We write it in the form U
\begin{alignat*}{1}
U&=(2k+1)^{-\frac12}\begin{pmatrix}
u_1(k+\frac12) & u_2(k+\frac12)\\
v_1(k)&v_2(k)
\end{pmatrix}\\
U^{-1}&=(2k+1)^{-\frac12}\begin{pmatrix}
v^1(k+\frac12) & u^1(k)\\
v^2(k+\frac12)&u^2(k)
\end{pmatrix}
\end{alignat*}
We have set the Dirac $b_\nu=u_\nu(k+\frac12)$ and $a^\nu=v^\nu(k+\frac12)$; since these matrices fulfill exactly 
the same relations which hold for $u_\nu(k), v^\nu(k)$ if $k$ is replaced by $k+\frac12$.
If $k$ passes through all the half-integers, we obtain a sequence of matrices with the following properties:

\begin{description}
\setlength{\itemindent}{.3in}
\item[$u_1(k),u_2(k)$] are rectangular matrices with $2k+1$ rows and $2k$ columns.
\item[$v_1(k),v_2(k)$] are rectangular matrices with $2k$ rows and $2k+1$ columns.
\end{description}
\marginnote{p.24}
In the sense of the matrix calculus, the following products can thus be formed:
\begin{description}
\setlength{\itemindent}{.3in}
\item[$v_\nu(k)u^\mu(k)$] Square matrix of rank $2k$.
\item[$u_\mu(k)v^\nu(k)$] Square matrix of rank $2k+1$.
\item[$v_\nu(k-\frac12)v^\mu(k)$] Rectangular matrix with $2k+1$ columns and $2k-1$ rows.
\item[$u^\mu(k)u^\nu(k-\frac12)$] Rectangular matrix with $2k-1$ columns and $2k+1$ rows.
\end{description} 
From the fact that the matrix $U$ constructed from the $u^\nu$ and $v^\mu$ brings the matrix $A(k)$
to a diagonal form, the equations follow:
\begin{alignat}{1}
u^\mu(k)v_\nu(k)&=\big(k\delta^\mu_\nu-s^\mu_{\cdot\,\nu}(k)\big)(-1)^{2k+1}\nonumber\\
v^\mu(k)u_\nu(k)&=\big(s^\mu_{\cdot\,\nu}(k-\tfrac12)+(k+\tfrac12)\delta^\mu_\nu\big)(-1)^{2k}\label{eq7.1}
\end{alignat}
Where $s^\mu_{\cdot\,\nu}(k)$ is the spinor associated with the infinitesimal rotations $\alpha^i(k)$.

Further 
\begin{alignat}{1}
v_\mu(\tfrac12)u^\nu(\tfrac12)&=\delta_\mu^\nu\nonumber\\
v^\mu(k)u_\nu(k)+u^\mu(k-\tfrac12)v_\nu(k-\tfrac12)&=2k\delta^\mu_\nu(-1)^{2k}\label{eq7.2}\\
u^\mu(k)v_\mu(k)=2k(-1)^{2k+1}\quad&;\qquad v_\mu(k)u^\mu(k)=(2k+1)(-1)^{2k+1}\nonumber
\end{alignat}
\begin{equation}\label{eq7.3}
u^\mu(k)u_\mu(k-\tfrac12)=v^\mu(k-\tfrac12)v_\mu(k).
\end{equation}
All these equations were derived by Dirac. But the $u_\nu, v_\nu$ are not considered as a function of $k$
(Dirac considers in particular $u^\nu(k), v_\nu(k)$ and $u^\nu(k+\tfrac12)=b^\nu, v^\nu(k+\tfrac12)=a^\nu$ for fixed $k$).

For our purposes the equations (\ref{eq6.3}) are especially important. They state that the rectangular matrices 
$u^\mu(k)u^\nu(k-\tfrac12)$ and $v^\mu(k-\tfrac12)v^\nu(k)$ in $\nu$ and $\mu$ are symmetric spinors. 
Now we construct the operators
\begin{alignat}{1}
R^{\mu_1\mu_2\dots\mu_{2k}}(k)&=v^{\mu_1}(\tfrac12)v^{\mu_2}(1)\dots v^{\mu_{2k}}(k)\nonumber\\
\text{and}\phantom{mmmmmmm}\hfill&\label{eq7.4}\\
P^{\nu_1\nu_2\dots\nu_{2k}}(k)&=u^{\nu_1}(k)u^{\nu_2}(k-\tfrac12)\dots u^{\nu_{2k}}(\tfrac12)\nonumber
\end{alignat}
$(R_l^{\mu\rho\dots\lambda})$ is symmetric in all spinor indices because of (\ref{eq6.3}). The index $l$ can assume 
$2k + 1$ values and, with respect to this index, $R^{\mu\rho\dots\lambda}$ can be regarded as a matrix with a single line.

$(P_n^{\delta\rho\dots\lambda})$ is also symmetric in all spinor indices. The index $n$ can assume $2k+1$ values, 
and $P^{\delta\rho\dots\lambda}$ can be considered as a matrix with a single column with respect to $n$.

\marginnote{p.25}
According to (6.2) and (6.4)
\begin{alignat}{1}
P^{\delta\rho\dots\lambda}_s(k)R_{r,\delta\rho\dots\lambda}&=(2k)\delta_{rs}\nonumber\\
\text{With the inverse formula}\phantom{mmmmmmm}\hfill&\label{eq7.5}\\
R_{s,\rho\delta\dots\lambda}(k)P_s^{\gamma\mu\dots\nu}(k)
&=\sum\mathrm{Perm}(\gamma\mu\dots\nu)\delta^\gamma_\rho\delta_\delta^\mu\delta_\lambda^\nu\nonumber
\end{alignat}
By means of the operator $P_s^{\gamma\mu\dots\nu}(k)$ one can now assign to each symmetric spinor $a_{\gamma\mu\nu}$ 
of rank $2k$ a quantity $A_s$ which has only one index passing through $2k + 1$ values. 
Conversely, one can assign a symmetric spinor to each $A_s$ by means of $R_{s,\rho\delta\dots\lambda}(k)$.
\begin{alignat}{1}
P_s^{\gamma\mu\dots\nu}(k)a_{\gamma\mu\dots\nu}&=\sqrt{(2k)!}A_s\nonumber\\
R_{s,\gamma\mu\dots\nu}(k)A_s&=\sqrt{(2k)!}a_{\gamma\mu\dots\nu}\label{eq7.6}
\end{alignat}
If the operation $P_s^{\gamma\mu\dots\nu}$ is applied to a non-symmetric spinor, this operation is no longer
unambiguously reversible according to (\ref{eq6.5}); rather, the symmetrized spinor is obtained. 
Therefore, $P_s^{\gamma\mu\dots\nu}$ can be used as a symmetrization operation.

Analog operators can also be defined for the dotted indices. We call them
\begin{equation}\label{eq7.7}
P^{\dot{s}}_{\dt\dm\dots\dn}(l),\qquad R^{\dot{s},\dt\dm\dots\dn}(l).
\end{equation}
They are constructed from the $u^{\dn}, v^{\dm}$ and satisfy the same equations as the operators with undotted indices, 
if in the formulas (\ref{eq7.1}) to (\ref{eq7.6}) every spinor $C_\mu$ is replaced by $C^{\dm}$, $b^\nu$ by $b_{\dn}$.

We can now proceed further by means of the products 
$$
v^\mu(k)v^\nu(k+\tfrac12)\dots v^\lambda(r)
$$
and
$$
u^\mu(k)u^\nu(k-\tfrac12)\dots u^\rho(l)
$$
transform only a part of the indices of a spinor into an index $S$, which gives a quantity which has indices of both kinds.
Conversely, such a quantity can again be transformed into a quantity $A^{\dot{s}}_r$. In particular, one can form 
variables which contain a spinor index.
This leads to the equations 
\begin{equation}\label{eq7.8}
\begin{cases}
p^{\dt\rho}\psi^A_\rho&=\kappa \psi^{\dt B}\\
p_{\dt\rho}\psi^{\dt B}&=\kappa\psi^A_\rho.
\end{cases}
\end{equation}
given by Dirac.

\marginnote{p.26}
Here
\begin{alignat*}{1}
\psi^{A\dot{r}}_{\rho,p}&=\sqrt{2k}v^s_{p,\rho}(k)A^{\dot{r}}_s\\
\psi^{B,\dt\dot{r}}_{p}&=\sqrt{2l}v^{\dot{r},\dt}_{\dot{t}}(l)B^{\dot{t}}_p\\
\end{alignat*}
$\psi^A_\rho,\psi^{B,\dt}$ satisfy the secondary conditions
\begin{equation}\label{eq7.9}
\begin{cases}
v_{\dt}(l-\tfrac12)p^{\dt\rho}\psi^A_\rho&=0\\
v^\rho(k-\tfrac12)p_{\dt\rho}\psi^{B\dt}&=0.
\end{cases}
\end{equation}
$A$ and $B$ therefore satisfy the equations 
\begin{alignat}{1}
p^{\dt\rho}v_\rho(k)A&=\sqrt{\frac{2l}{2k}}\kappa v^{\dt}(l)B\nonumber\\
\nonumber\\
p_{\dt\rho}v^{\dt}(l)B&=\sqrt{\frac{2k}{2l}}\kappa v_\rho(k)A\label{eq7.9} 
\end{alignat}
as well as the secondary conditions: 
\begin{equation}\label{eq7.10}
v_{\dt}(l-\tfrac12)p^{\dt\rho}v_\rho(k)A=v^\rho(k-\tfrac12)p_{\dt\rho}v^{\dt}(l)B=0.
\end{equation}

\subsection{Quantization of field theories for half-integer spin $f$}
Again, as in the case of integer spins, we want to establish the commutation relations between the field magnitudes, 
which are identical for all equations to which the field magnitudes satisfy.

In order to establish such relations, one proceeds from those which satisfy the above-defined spinor 
$a^{(m)}_{\alpha\beta\dots\gamma}$, which has only undotted indices. The number of indices is the odd number $2m + 1$.
 $a^{(m)}_{\alpha\beta\dots\gamma}$ is symmetric in all indices, while the secondary condition (\ref{eq5.3}) is 
 eliminated, which leads to a certain simplification.

As a commutation relation between $a^{(m)}$ and $a^{*(m)}$ we assume: 
\begin{equation}\label{eq8.1}
\frac1i[a^{(m)}_{\alpha\beta\dots\gamma},a^{(m)*}_{\dn\dr\dots\dl}]^+
=\frac{1}{\kappa^{2m}(2m+1)!}\sum\mathrm{Perm}(\alpha\dots\gamma)p_{\alpha\dt}\dots p_{\gamma\dn}D(x).
\end{equation}
It means $[a, b]^+=[ab+ba]$. $D (x)$ is again the invariant function defined by (\ref{eq4.3b}). 
On the right-hand side of (\ref{eq7.1}) there is an odd number, namely $2m+1$ differentiations $p_{\dr\delta}$.

\marginnote{p.27}
The relations (\ref{eq8.1}), if they can be fulfilled at all, result in the so-described particles 
fulfilling the Pauli exclusion principle; a circumstance which makes it possible to make the energy 
positive by means of a "hole theory." We can show that the relations are satisfiable, as follows:

Let $k_{\dl\beta}$ be the spinor associated with the wave vector $k_i$ according to $k_{\dl\beta}=k_i\sigma^i_{\dl\beta}$.
Then, if the definition of $D(x)$ is observed, the commutation relations in the momentum space are 
\begin{equation}\label{eq8.2}
\frac1i[a_{\alpha\beta\dots}(k),a^*_{\dn\dr}(k')]^+=\frac{\delta_{kk'}}{2\kappa^{2m}(2m+1)!}
\sum\mathrm{Perm}(\alpha\beta\dots)k_{\alpha\dn}k_{\beta\dr}\dots\frac1{k_4}.
\end{equation}
From this, it can be seen that waves belonging to different $k_i$ give the plus-commutation zero. 
Let us consider two waves belonging to the same $k_i$, and indeed in their rest frame. The right-hand side 
of (\ref{eq8.2}) is therefore not equal to zero if $a_{\alpha\beta\dots\gamma}$ is the conjugate-complex of $a^*_{\dn\dr\dots\dl}$. The left side is therefore never negative, so must also be the right one.
The right side is now either zero or equal to 
\begin{equation}\label{eq8.3}
\frac1{2\kappa^{2m}}\binom{2m+1}{s}^{-1}\left(\frac{k_4}{i}\right)^{2m}=\frac12\binom{2m+1}{s}^{-1}
\end{equation}
which is positive. ($s$ is the number of indices of $a_{\alpha\beta\dots}(k)$ which are equal to 1.) The commutation
relation therefore has the desired form 
$$
[a_i,a^*_k]=\delta_{ik}\cdot\text{const}.
$$
That the charge has the proper eigenvalues on the basis of the commutation relations (\ref{eq8.2}) follows immediately 
if we consider the charge of a plane wave in the rest frame. It has the positive definite form $\sum aa^*$ 
(according to (\ref{eq6.2}), where each term is just $2\binom{2m+1}{s}$ times, which factor according to (\ref{eq8.3})
is just compensated by our commutation relations.

If the spin is an integer, we can also write the commutation relations between the quantities 
$a^{(b)}$ defined in Section 3 in the momentum space in the form 
\begin{equation}\label{eq8.4}
\frac1i[a^f_{\alpha\beta\dots\gamma}(k),a^{f*}_{\dn\dr\dots\dl}(k)]^-
=C\delta_{kk'}\sum\mathrm{Perm}(\dn\dr\dots)k_{\alpha\dn}\dots\frac{1}{k_4}.
\end{equation}
Here, however, an even number of factors $k_{\dl\delta}$ occur, namely $2f$.
\marginnote{p.28}
Therefore, in the rest frame, the right-hand side of (\ref{eq8.4}) has the form
$$
C\left(\frac{k_4}{i}\right)^{2f-1}
$$
This is positive or negative depending on $\frac{k_4}i =\pm \kappa$. Therefore, the left-hand side of (\ref{8.4}) can not be
written as a commutation relations with the plus sign, since the left-hand side would assume the positive-definite form 
$$
[a^*a+aa^*]
$$
while the right-hand side could be positive or negative. 
Thus, particles with integer spin can not be quantized according to the exclusion principle without renouncing 
the infinitesimal character of the commutation relations, but only according to 
Bose statistics.\footnote{See: W. Pauli, Annales Poincaré VI (1936) p. 147 \& sq.}
Particles with half-integer spin can and must be quantized according to the exclusion principle, so that the energy 
becomes positive. From the above perspective, the long-assumed relationship between spin and statistics seems to be 
mathematically proved in a simple way. Moreover, it is immaterial that the spinors are irreducible. It is only the existence of 
a rest frame for each plane wave, the properties of the $D$-function, and the fact that the number of indices is even or odd, 
according to which the spin is a whole or half-integer. The occurrence of the $D$ function means that the commutation relations 
should be relativistic invariant and infinitesimal.

From the inversion relations (\ref{eq8.1}) for the $a^{(m)}_{\alpha\beta}$ one can obtain for arbitrary $a^{(q)}$ and 
$a^{(q)*}$, which have dotted and undotted indices, by applying the operation $\frac{p^{\dl\beta}}\kappa$ to (\ref{eq7.1}) according to
\begin{alignat*}{1}
\frac1i[&a^{(m-1)\dn}_{\alpha\beta\dots\gamma},a^{(m-1)*\beta}_{\dl\dr\dots\dm}]
\frac1{i\kappa^2}[p^{\dn\delta(m)}a_{\delta\alpha\dots\gamma},p^{\dt\beta(m)*}a_{\dt\dl\dots\dm}]\\
&=p^{\dn\delta}p^{\dt\beta}\frac1{n^{2m+2}(2m+1)!}
\sum\mathrm{Perm}(\alpha\beta\dots)p_{\delta\dt}p_{\alpha\dl}\dots p_{\gamma\dm}D(x).
\end{alignat*}
We obtain, in this way, the commutation relations for the spinor $a^{(m-n)\dn\dots}_{\alpha\dots}$ of the $2m + 1-n = p$ 
undotted and $n$ dotted indices with its conjugate
\begin{alignat}{1}
\frac1i[&a^{(m-n)\dl_1\dots\dl_n}_{\mu_1\dots\mu_p},a^{(m-n)*\dl_1\dots\dl_n}_{\dm_1\dots\dm_p}]
=\frac{\kappa^{-2m}}{(2m+1)!}\sum_{l=0}^n\sum\pi_l \label{eq8.5}\\
&p_{\mu_1\dm_1\dots}p_{\mu_{p-l}\dm_1{p-l}}\delta^{\lambda_1}_{\mu_{p-l+1}}\delta^{\dl_1}_{\dm_1{p-l+1}}\dots
\delta^{\dl_l}_{\dm_p}p^{\mu_{l+1}\dm_{l+1}}\dots p^{\lambda_n\dl_n}D(x)\nonumber
\end{alignat}

\marginnote{p.29}
$\sum\pi_l$ means the permutation operator of the indices of the subsequent summator, which symmetrizes 
the term with as few commutations as possible.
The number of all summands is then just $(2m+1)!$ as it must be, since by the differentiation this number
is not changed. By the operators $P^s$ defined by (\ref{eq7.4}) to (\ref{eq7.6}) we can also write our 
relations in the following form: 
\begin{alignat}{1}
\frac{n!p!}{i}& [A^{(m-n)\dot{s}}_t, A^{(m-n)*q}_{\dot{r}}]
=\frac{\kappa^{-2m}}{(2m+1)!}P^{\dot{s}}_{\lambda_1\dots\lambda_n}P^{*\dot{q}}_{\dl_1\dots\dl_n}
P^{\mu_1\dots\mu_p}_tP^{*\dm_1\dots\dm_p}_{\dot{r}}\sum_{l=0}^p\frac1{l!(p-l)!}\cdot\nonumber\\
&\frac{1}{l!(n-l)!}\kappa^{2l}p_{\mu_1\dm_1}\dots p_{\mu_{p-l}\dm_{p-l}}
\delta^{\dl_1}_{\dm_{p-l+1}}\delta^{\dl_l}_{\dm_p}\dots p^{\lambda_{l+1}\dl_{l+1}}\dots p^{\lambda_n\dl_n}
D(x).\label{eq8.6}
\end{alignat}

The commutation relations between the variables $a^{(q)}$ and $b^{(q)*}$ can also be indicated in a 
similarly simple manner. (When mirrored, $a^{(q)}$ and $b^{(q)*}$ are exchanged.)
One finds
\begin{alignat}{1}
\frac{1}{i} [a^{(q)\dl_1\dots\dl_r}_{\rho_1\dots\rho_s}, a^{(q)*\dn_1\dots\dn_s}_{\dm_1\dots\dm_r}]
=\frac{1}{(r+s)!}\sum_{k=0}^r&\pi_k\kappa^{(1-2k)}p^{\dl_1\nu_1}p^{\dl_2\nu_2}\dots p^{\dl_k\nu_k}p_{\dm_1\rho_1}\dots p_{\dm_k\rho_k}\cdot\nonumber\\
&\delta^{\nu_{k+1}}_{\rho_{k+1}}\dots\delta^{\nu_s}_{\rho_s}\delta^{\dl_{k+1}}_{\dm_{k+1}}\dots\delta^{\dl_r}_{\dm_r}D(x).\label{eq8.7}
\end{alignat}
Where $\sum\pi_k$ is the same permutation operator as $\sum\pi_l$ in (8.5).

\appendix
\renewcommand\theequation{\arabic{equation}}
\section*{Appendix}
\tocadd{Appendix}
\subsubsection*{Determination of the coefficients of the formula \ref{eq8.5}}
\setcounter{equation}{0}
\subtocadd{Determination of the coefficients of the formula (\ref{eq8.5})}
We look at an expression of the following form
\begin{equation}\label{1}
p_{\mu_1\dot{\mu}_1}\dots p_{\mu_r\dot{\mu}_r}\delta^{\lambda_1}_{\mu_{r+1}}\delta^{\dot{\lambda}_1}_{\dot{\mu}_{r+1}}
p^{\lambda_{n+1}\dot{\lambda}_{n+1}}\dots p^{\lambda_{k+n}\dot{\lambda}_{k+n}}
\end{equation}
where
$$
\begin{aligned}
r+n&=p\\
k+n&=q\qquad\text{as well as}\quad p+q=L.
\end{aligned}
$$
(\ref{1}) must now be symmetrized into the $\mu_i,\dot{\mu}_i,\lambda_k,\dot{\lambda}_k$. We thus have to form a sum
of terms of the form (\ref{1}), where the $\mu_i,\lambda_i,\dots$ are permuted appropriately. How many terms must at least 
contain the sum so that it is symmetric in the indices $\mu_i,\lambda_i,\dots$?

We decide to do so: We first symmetrize from the $\mu_1$ to $\mu_r$. What emerges is symmetrical in the $\dot{\mu}_1$ to 
$\dot{\mu}_r$ by itself. We also symmetrize the $\lambda_{n+1}$ to $\lambda_q$. Thus we obtain $k!r!$ terms. 
Now we symmetrize the $\lambda_1$ to $\lambda_n$ as well as the $\dot{\lambda}_1$ to $\dot{\lambda}_n$, 
whereby the $\mu_{r+1}$ to $\mu_{r+n}$ as well as the $\dot{\mu}_{r+1}$, to $\dot{\mu}_p$ are also 
symmetrized. This gives 
$$
(n!)^2k!r!\quad\text{terms.}
$$

\marginnote{p.30}
Now one can divide $\lambda_1$ to $\lambda_q$ into three groups with $k$ and $n$ indices in $\binom{q}{n}$ ways. 
The same holds for the dotted $\dot{\lambda}_1$ as for the $\mu$, with respect to $r$ and $n$. We thus obtain 
\begin{equation}\label{2}
(n!)^2r!k!\binom{p}{n}^2\binom{q}{n}^2=p!q!\binom{p}{n}\binom{q}{n} \quad\text{terms.}
\end{equation}
Then the resulting sum is completely symmetrical. Let's call this sum $S_n$. The symmetrization operation generated by (\ref{1}) $S_n$ is called $\pi_n$. If we choose from the sum: 
\begin{equation}\label{3}
\sum\mathrm{Perm}(\delta_i)p_{\mu_1\dot{\mu}_1}\dots p_{\mu_2\dot{\mu}_2},
\end{equation}
which contains $L!$ terms and then applies the operation 
\begin{equation}\label{4}
\frac1{p^2}p^{\lambda_k\dot{\mu}_i}p^{\dot{\lambda}_k\mu_i}
\end{equation}
$q$ times, we arrive at a sum symmetrical in the $\mu_i,\dot{\mu}_i$ and $\lambda_k,\dot{\lambda}_k$ of the form 
\begin{equation}\label{5}
\sum_nS_nC_n.
\end{equation}

It is easy to conceive that due to the origin of (\ref{5}), all $C_n$ must be non-zero natural numbers.
We now show that they are all equal. The sum (\ref{3}) contains $L!$ terms, which are all multiplied by 
a factor of one. The operation (\ref{4}) does not change this number, and can not give rise to any number 
of factors. Consequently, the sum (\ref{5}) must also contain $L!$ summands. 
Since $S_n$ contains $p!q!\binom{p}{n}\binom{q}{n}$ summands, then it must give
$$
\sum_nC_np!q!\binom{p}{n}\binom{q}{n}=L!
$$
Now $\sum(p_n)(q_n)$ is the constant term of
$$
(1+x)^p(1+\frac1x)^q=\frac{(1+x)^L}{x^q}.
$$
But this is $\binom{L}{q}$, from which it follows that $\sum_n p!q!\binom{p}{n}\binom{q}{n}=L!$, so that all $C_n = 1$. 
If an expression of the form (\ref{1}) is symmetrized by the $u_\nu, u_{\dot{\nu}}$, then this means that all the 
permutations of the $\mu_i,\dot{\mu}_i$ and the $\lambda_i,\dot{\lambda}_i$ are summed. Then one obtains 
$(p!)^2(q!)^2$ summands and the result is 
$$
(u!)^2(q-n)!^2(p-n)!^2 S_n
$$
which is why, by symmetrization by means of the $u_\nu,u_{\dot{\nu}}$, the factor 
$\frac1{(n!)^2(q-n)!(p-n)!}$ must be set before each summand of the form (\ref{1}).

\newcommand{\ua}{\underline{a}}
\setcounter{equation}{0}
\marginnote{p.31}
\subsubsection*{Number of independent components of tensors}
\subtocadd{Number of independent components of tensors}
A tensor of rank $f$ in the $k$-dimensional space which is symmetric in all indices has 
$$
\binom{f+k-1}{k-1} \quad\text{components.}
$$
Our tensor $A_{ik\dots l}$ of rank $f$ is symmetric in all indices and satisfies equations
$$
A_{ii\dots k}=0.
$$
This says that a symmetric tensor of rank $f - 2$ is to disappear, that is, there are as many equations as such a tensor component. In our case the space has 4 dimensions. $A_{ik\dots l}$ therefore has $\binom{f+3}{3}$ components between which there are $\binom{f+3}{3}$ equations.

Therefore, $A_{ik\dots l}$ has
$$
\binom{f+3}{3}-\binom{f+1}{3}=(f+1)^2
$$
linearly independent components.

Because of the differential equation 
$$
\frac{\partial A_{ik\dots l}}{\partial x_i}=0
$$
the amplitudes are zero for plane waves in their rest frame if $i, k\dots = 4$. The dimension $k$ is therefore reduced
to 3 in this case. Therefore, there exist 
$$
\binom{f+2}{2}-\binom{f}{2}=2f+1
$$
linearly independent planar waves on given waves.

We now want to determine the number of components of the tensors $B^{(1)}_{\underline{a}k(ml)}$ ($\underline{a}$ means the set of $f-2$ indices. In the indices $(\underline{a}k)$, $B^{(1)}$ is symmetric, skew in $[ml]$. $B^{(1)}$ satisfies the equation 
$$
\begin{aligned}
B^{(1)}_{\underline{a}k[kl]}&=0.\\
\\
B^{(1)}_{\underline{a}k[lm]}+B^{(1)}_{\underline{a}m[kl]}+B^{(1)}_{\underline{a}l[mk]}&=0.
\end{aligned}
$$
Now the independence of these equations must be determined. 
\marginnote{p.32}
To this end, we write out the equations in abbreviated form by writing only the indices. We continue with 

\begin{alignat*}{1}
[23]=[1]\qquad &[14]=[4]\\
[31]=[2]\qquad &[24]=[5]\\
[12]=[3]\qquad &[34]=[6].
\end{alignat*}
This gives us
\begin{equation}\label{a1}
\begin{cases}
\ua2[3]-\ua3[2]-\ua4[4]&=0\\
\ua1[3]-\ua3[1]-\ua4[5]&=0\\
\ua2[1]-\ua1[2]-\ua4[6]&=0\\
\ua1[4]+\ua2[5]+\ua3[6]&=0\ (+)\\
\end{cases}
\end{equation}
\begin{equation}\label{a2}
\begin{cases}
\ua3[5]-\ua2[6]-\ua4[1]&=0\\
\ua1[6]-\ua3[4]-\ua4[2]&=0\\
\ua2[4]-\ua1[5]-\ua4[3]&=0\\
\ua1[1]+\ua2[2]+\ua3[4]&=0\ (+)\\
\end{cases}
\end{equation}
The last two equations in (\ref{a1}) and (\ref{a2}) which are denoted by $(+)$ can be deduced from the three others 
by addition if an index among the indices $a$ is equal to 4. Then, from (\ref{a1}) and (\ref{a2}), by substituting the equations 
$$
B_{ii\ua[ml]} = 0 
$$
in each equation for $\ua$, which therefore means nothing new. Otherwise, (\ref{a1}) and (\ref{a2}) are independent equations.

$B^{(1)}_{ik\dots[ml]}$ is now symmetric in $f-1$ indices and skewed in two, therefore it has $6\binom{f+2}{3}$ components.
Further, it suffices (\ref{a1}) and (\ref{a2}). In six of these equations the $(f- 2)$ indices $\ua$ run from 1 to 4. 
These equations therefore mean that a symmetric tensor of rank $f-2$ should be zero in four dimensions. 
The equations therefore give $6\binom{f+1}{3}$ ancillary conditions. 
However, in the last two equations denoted by $(+)$, the indices a only run from 1 to 3, 
if only independent equations are considered. 
We therefore obtain $2\binom{f}{2}$ further conditions. Accordingly, $B^{(1)}$ has
$$
6\binom{f+2}{3}-6\binom{f+1}{3}-2\binom{f}{2}=2f^2+4f
$$
linearly independent components.

\marginnote{p.33}
\setcounter{equation}{0}
\subsubsection*{Fields with null rest mass}
\subtocadd{Fields with null rest mass}
To consider the case of the rest-mass zero, let us assume the following equations:

In the case of whole spins:

\begin{equation}
\begin{cases}
\square A_{ik\dots l}&=0\\
\frac{\partial A_{ik\dots l}}{\partial x_i}&=0
\end{cases}
\end{equation}
It follows that:
\begin{equation}
\begin{cases}
\frac{\partial A_{ik\dots l}}{\partial x_m}-\frac{\partial A_{mk\dots l}}{\partial x_i}&=B^{(1)}_{[mi]k\dots l}\\
\frac{\partial B_{[mi]k\dots l}}{\partial x_m}&=0.
\end{cases}
\end{equation}

In the case of half-integer spins:
\begin{equation}
\begin{cases}
p ^{\dot{\nu}\rho}a^{(0)\dot{\lambda}\dots}_{\rho\delta\dots}&=0\\
p_{\dot{\lambda}\lambda}b_{\rho\dots}^{\dot{\lambda}\dot{\mu}\dots}&=0.
\end{cases}
\end{equation}

From these equations follows that: 
\begin{equation}
\begin{cases}
\square a^{(0)\dot{\lambda}\dots}_{\rho\delta\dots}&=0\\
\square b_{\rho\dots}^{(0)\dot{\lambda}\dot{\mu}\dots}&=0\\
\end{cases}
\end{equation}

These equations arise from those which we have given above by nullifying $\kappa$. The 
$A_{ik\dots l},a^{(0)\dot{\lambda}\dots}_{\rho\delta\dots},b_{\rho\dots}^{(0)\dot{\lambda}\dot{\mu}\dots}$  are otherwise 
to have the same algebraic properties as in the case of $\kappa\not=0$. By means of these field magnitudes we can 
again construct an  energy-momentum tensor and a current vector. The energy tensor in the integer case is 
\begin{equation}
T_{kl}=\frac12\{B^{*(1)}_{[ik]m\dots}B^{(1)}_{[il]m\dots} + B^{(1)*}_{[il]m\dots}B^{(1)}_{[ik]m\dots} \}
-\frac14\delta_{kl}B^{*(1)}_{[ri]m\dots}B^{(1)}_{[ri]m\dots}
\end{equation}
The energy
\begin{equation}
-\int T_{44}dN=\sum_k|k_4|^2\{A^{*+}_{ir}(k)A^+_{ir}(k)+A^{-*}_{ir}(k)A^-_{ir}(k)\}
\end{equation}
In the half-integer case, the current vector is
\begin{equation}
s_{\dot{\lambda}\beta}=
a^{(0)*\rho\dots}_{\dot{\lambda}\dot{\nu}\dots}a^{(0)\dot{\nu}\dots}_{\beta\rho\dots}
+
b_{\beta\dot{\mu}\dots}^{(0)*\rho\dots}b_{\dot{\lambda}\rho\dots}^{(0)\dot{\mu}\dots}.
\end{equation}
With the total charge
\begin{equation}
\rho=\int(s_{1\dot{1}}+s_{2\dot{2}})dV.
\end{equation}
\marginnote{p.34}
The expressions (6) and (8) can vanish for certain states without the field magnitudes being zero. 
In the case of the integer spin the energy is zero whenever the tensor $A_{ik\dots l}$ can be represented as follows: 
\begin{equation}
N_{ik\dots lm}=\frac{\partial C_{k\dots lm}}{\partial x_i}+\frac{\partial C_{i\dots lm}}{\partial x_k}\dots+\frac{\partial C_{ik\dots l}}{\partial x_m}
\end{equation}
where $C_{k\dots lm}$ satisfies the following equations: 
$$
\square C_{k\dots l}=0,\quad \frac{\partial C_{k\dots l}}{\partial x_k}=0,\quad C_{kk\dots l}=0
$$
and is symmetric in all indices. The simplest way to show that the energy is zero is to decompose $C_{k\dots l}$ into plane waves and use the expression for the energy.

It is also possible to add a field $N_{ik\dots l}$ of the form defined by (9)  to any field 
$A_{ik\dots l}$, without changing the energy. In analogy to electrodynamics, the transformation 
$$
A'_{ik\dots l}=A_{ik\dots l}+N_{ik\dots l}
$$
is called "gauge transformation". So $N_{ik\dots l}$ is defined by (9).

If $A_{ik\dots l}$ has $f$ indices, then there are $2f + 1$ linearly independent plane waves $A_{ik\dots l}$
for each wave vector. The tensor $C_{ik\dots l}$ has $f-1$ indices, Accordingly, there are $2f-1$ linearly independent plane 
waves $N_{ik\dots l}$ for every wave vector.  If states which are separated by "gauge transformation" are regarded 
as equivalent then there are only two linearly independent, truly different plane waves of given wave number and frequency at rest mass zero and $f\geqslant 1$.

In the same way one can add to the fields $a^{(0)\dot{\lambda}\dots}_{\beta\dots}, b^{(0)\dot{\nu}\dots}_{\delta\dots}$ spin-field $n^{(0)\dot{\lambda}\dots}_{\beta\dots},m^{(0)\dot{\nu}\dots}_{\delta\dots}$ without changing the energy and the charge. Here, $n^{(0)\dot{\lambda}\dots}_{\beta\dots},m^{(0)\dot{\nu}\dots}_{\delta\dots}$ are of the form 
\begin{alignat*}{1}
n^{\dl\dt\dots}_{\gamma\delta\dots}&=p^{\dl}_{\gamma}c^{\dt\dots}_{\delta\dots}
+p^{\dt}_\gamma c^{\dl}_{\delta}+\dots+p^{\dl}_\delta c^{\dt}_\gamma+\dots\\
m^{\dl\dt\dots}_{\gamma\delta\dots}&=p^{\dl}_{\gamma}d^{\dt\dots}_{\delta\dots}+p^{\dt}_\gamma d^{\dt}_{\delta}+\dots+p^{\dl}_\delta d^{\dt}_\gamma+\dots
\end{alignat*}
$C^{\dot{\tau}}_\delta$ has in each case a dotted and an undotted index less than $a^{(0)}, \alpha^{\dn}_\delta$ a dotted and an undotted index less than $b^{(0)}$. Both are symmetric and satisfy the same equations as $a^{(0)}$ and $b^{(0)}$, respectively.
\marginnote{p.35}
It follows from this, as in the case of integer spins, if two states of the wave number and the frequency are not really different, then there are only two really different plane waves of states, which differ from one another by the "gauge transformation"
\begin{alignat*}{1}
a^{(0)\dot{\lambda}\dots}_{\gamma\dots}+n^{\dot{\lambda}\dots}_{\gamma\dots}&=a^{(0)'\dot{\lambda}\dots}_{\gamma\dots}\\
b^{(0)\dot{\lambda}\dots}_{\gamma\dots}+m^{\dot{\lambda}\dots}_{\gamma\dots}&=b^{(0)'\dot{\lambda}\dots}_{\gamma\dots}
\end{alignat*} 
which therefore belong to the same energy and charge.(In the case of spin $\frac12$, of course, the gauge transformation is lost.)

From the $B^{(1)}$ and the $a^{(0)}, b^{(0)}$ one can again form new quantities $B^{(q)}, a^{(q)}, b^{(q)}$. 
With these, however, no physically usable energy tensors can be constructed because the corresponding energy vanishes 
identically in accordance with the wave equations. Interestingly, however, those spinors which have only a single
index location and are obtained by differentiation from the spinor associated with $A_{ik\dots l}$, or from $a^{(0)}, b^{(0)}$
by differentiation according to 
\begin{equation}
a_{\alpha\beta\dots\gamma\rho}=p_{\gamma\dm}p_{\rho\dn}\dots a^{(0)\dm\dn\dots}_{\alpha\beta\dots}.
\end{equation}
These quantities are "gauge-invariant" and vanish when charge and energy disappear. So the theory can be quantified 
with rest mass zero by simply asking the following interconversion relations between these gauge-invariant field magnitudes, 
\begin{equation}
\frac1i[a^*_{\alpha_1\dots\alpha_k}, a_{\dt_1\dots\dt_k}]^{\pm}
=\frac1{k!}\sum\mathrm{Perm}(a_i)p_{\alpha_1\dt_1}\dots p_{\alpha_k\dt_k}D(x)
\end{equation}
taking the + or - sign as the case may be $k$ is odd or even. Since the theory is rotationally invariant, and all Fourier 
components are commutable, the contradiction of the commutation relations can easily be viewed by considering a plane wave in 
the $z$-direction. Let us, however, dispense with the very simple proofs. According to (10), all other field magnitudes can be 
recovered by the integration from the "gauge-invariant" field quantities, with the exception of a gauge transformation, in 
particular the quantities $a^{(0)}, b^{(0)}$ forming the energy tensor and the tensor $A_{ik\dots l}$.
Since the total energy and the total charge are gauge-invariant, then the commutation relations (10) are sufficient to determine 
the eigenvalues of these quantities.
\marginnote{p.36}
Since the energy and charge density are constructed from $a^{(0)}, b^{(0)}$ and $B^{(1)}_{[ik]l\dots}$, it follows 
that these quantities can not be invariable except for spin $0, \frac12$ and $1$, which means that the small spin 
values are excellent

\subsubsection*{Special cases as examples}
\subtocadd{Special cases as examples}

We shall briefly summarize here a few formulas of the theory developed here, which refer to the special cases $f = \frac32$ and $f = 2$. For the case $f = 1$, we refer to the literature cited.

\begin{itemize}
\item[\underline{$f=\frac32$.}]Here are the equations in Spinor form 
\begin{alignat*}{1}
p^{\dl\mu}a^{\dt}_{\mu\lambda}=\kappa b^{\dl\dt}_\lambda;\quad&\epsilon_{\dl\dt}p^{\dt\rho}a^{\dl}_{\rho\lambda}=0\\
\\
p_{\mu\dl}b^{\dl\dt}_{\lambda}=\kappa a^{\dt}_{\mu\lambda};\quad&\epsilon^{\mu\lambda}p_{\mu\dl}b^{\dl\dt}_{\lambda}=0.
\end{alignat*}
The commutation relations take the form
\begin{alignat*}{1}
\frac1i[a^{*\mu}_{\dl\dt},a^{\dm}_{\alpha\beta}]^+=\frac16&[p_{\dl\alpha}\delta^{\dm}_{\dt}\delta^\mu_\beta+p_{\dt\alpha}\delta^{\dm}_{\dl}\delta^\mu_\beta+p_{\dt\beta}\delta^{\dm}_{\lambda}\delta^\mu_\alpha\\
&+\frac1{\kappa^2}\{p_{\dl\alpha}p_{\dt\beta}p^{\dm\mu}+p_{\dl\beta}p_{\dt\alpha}p^{\dm\mu}\}]D(x).
\end{alignat*}
Looking at plane waves that change as $e^{ikz+i\omega t}$ in space and time, the field equations represent the existence of the relations
\begin{alignat*}{1}
k^2-\omega^2&+\kappa^2=0\\
\\
a^{\dot{1}}_{21}=a^{\dot{2}}_{11}\frac{\omega+k}{\omega-k};\quad&a^{\dot{2}}_{12}=a^{\dot{1}}_{22}\frac{\omega-k}{\omega+k}
\end{alignat*}
$a^{\dot{1}}_{22},a^{\dot{1}}_{11},a^{\dot{2}}_{22},a^{\dot{2}}_{11}$ can then be considered as the four independent amplitudes which denote the four polarizations.

\item[\underline{$f=2$.}] Here we would like to state only the commutation relations between the $A_{ik}$:
$$
i[A_{ik}, A^*_{i'k'}]=\frac12(R_{ii'}R_{kk'}+R_{ik'}R_{ki'}-\frac23 R_{ik}R_{i'k'})D(x).
$$
The factor $\frac12$ of the right-hand side is chosen so that the track over $i, i ', k, k'$ on the right-hand side becomes equal to $(2f+1)D=5D$. (Note that $R_{ik}R_{kl} = R_{il}$; $R_{kk} = 3$).
\end{itemize}
\marginnote{p.37}
\subsubsection*{Special representation of the $u^\nu$ and $v^\nu$.}
\subtocadd{Special representation of the $u^\nu$ and $v^\nu$.}
Let $\alpha$ be an index which indicates how many indices of a spinor are one and which runs from zero to $2k-1$, $\beta$
 is a plane index running from zero to $2k$. If $k$ is an integer, then set
\begin{alignat*}{3}
u_1(k)_{\beta\alpha}&=\sqrt{\beta}\delta_{\beta-1,\alpha},\qquad& v^1_{\alpha\beta}(k)&=\sqrt{\alpha+1}\delta_{\alpha+1,\beta}\\
u_2(k)_{\beta\alpha}&=\sqrt{2k-\beta}\delta_{\beta\alpha},\qquad& v^2_{\alpha\beta}(k)&=\sqrt{2k-\alpha}\delta_{\alpha\beta}
\end{alignat*}
If $k$ half-integer, we put
\begin{alignat*}{3}
u^1_{\beta\alpha}(k)&=\sqrt{\beta}\delta_{\beta-1,\alpha},\qquad& v_{1,\alpha\beta}(k)&=\sqrt{\alpha+1}\delta_{\alpha+1,\beta}\\
u^2_{\beta\alpha}(k)&=\sqrt{2k-\beta}\delta_{\beta\alpha},\qquad& v_{2,\alpha\beta}(k)&=\sqrt{2k-\alpha}\delta_{\alpha\beta}.
\end{alignat*}
The representation of the Lorentz group resulting from this representation according to
$$
\mu^\mu(k)v_\nu(k)=(k\delta^\mu_\nu-s^\mu_\nu(k))(-1)^{2k+1}
$$
is identical with that of van der Waerden.

The following is given in the designation of van der Waerden
$$
A_p=s^{11},\quad A_q=s^{22},\quad A_2=s^{12}
$$
and
$$
J=k,\quad J+M=\beta.
$$
This work was carried out under the direction of Prof. W. Pauli; I would like to thank him for many valuable suggestions.

\hfill Zürich, Physics Institute of the E.T.H.

\appendix
\section{Remarks}
This translation has been made available, as is, for whoever is interested, even though it was made for the personal 
use of the author. If you find any mistake or better expressions, feel free to contact the author of the translation for an update.

The structure of the paper is respected as faithfully as possible. The references are those of the original paper, as 
for the notations and the equations numbering. The original paper did not contain a table of contents though.

The page numbers in the margin refer to the original page numbering.

The original German paper is currently available at 
\vskip1pc

\url{http://www.e-periodica.ch/digbib/view?pid=hpa-001:1939:12::13#13}


\begin{thebibliography}{9}
\bibitem{jordan:pauli}
    P. Jordan and W. Pauli, Z. S. f. Ph. {\bf 47} (1928), S. 151
\bibitem{waerden}
    Van der Waerden, Die Gruppentheoret. Methode in der Quantenmechanik, Springer, Berlin 1932. III. Kapitel, § 20.\TN{An English translation of this paper is available on  \url{https://arxiv.org/abs/1703.09761}.}
\bibitem{dirac}
    P.A.M. Dirac, Proc. Roy. Soc. {\bf A 155} (1936), S. 447.
\bibitem{sakata:yukawa}
    S. Sakata und H. Yukawa, Proc. Phys.-Math. Soc. Japan {\bf 19} (1937), S. 91.
\bibitem{proca}
    Proca, Comptes Rendus {\bf 202} (1936), S. 1490.
\bibitem{stuckelberg}
    E.C.G. Stückelberg, Helv. Ph. Acta XI. (1938), S. 299.
\bibitem{kemmer}
   N. Kemmer, Proc. Roy. Soc. {\bf A 166} (1938), S. 127. Here we can also find applications of this theory to nuclear forces as well to the hard component of cosmic rays.
\end{thebibliography}
\end{document}